# Credit, Land Speculation, and Long-Run Economic Growth

Tomohiro Hirano and Joseph E. Stiglitz[*]

**Abstract**

This paper presents a model that studies the impact of credit expansions arising from increases in collateral values or lower interest rate policies on long-run productivity and economic growth in a two-sector endogenous growth economy, with the driver of growth lying in one sector (manufacturing) but not in the other (real estate). We show that it is not so much aggregate credit expansion that matters for long-run productivity and economic growth but sectoral credit expansions. Credit expansions associated mainly with relaxation of real estate financing (capital investment financing) will be productivity-and growth-retarding (enhancing). Without financial regulations, low interest rates and more expansionary monetary policy may so encourage land speculation using leverage that productive capital investment and economic growth are decreased. Unlike in standard macroeconomic models, in ours, the equilibrium price of land will be finite even if the safe rate of interest is less than the rate of output growth.

[*] First version, April 19th 2021, This version, May 5th 2024. We thank seminar participants at various institutions and conferences for comments. Tomohiro Hirano (Royal Holloway, University of London), tomohih@gmail.com; Joseph E. Stiglitz (Columbia University), jes322@columbia.edu Stiglitz gratefully acknowledges financial support from the Hewlett and Sloan Foundations.

**Section 1. Introduction**

Financial liberalization and expansionary monetary policy have been widely supported as leading to enhanced economic growth. The experience has been otherwise. Verner (2019) showed that rapid expansion in credit to households and firms systematically predicts growth slowdowns.[1] The credit expansion of the early part of this century suggests an explanation: credit went disproportionately into real estate—so much so that the associated real estate boom crowded out more productive investments in other sectors. Müller and Verner (2023) confirm that this is typical: using a novel database on the sectoral distribution of private credit for 117 countries starting in 1940, they show that credit expansions to the construction and real estate industries systematically predict subsequent productivity and growth slowdowns, while credit expansions to the manufacturing sector are associated with higher productivity and long-run growth.[2]

This paper presents a model that is consistent with these empirical results. Key is endogenous growth, originating in only one sector ("manufacturing") but with spillovers to the other ("real estate"). In such situations, there is no presumption for the optimality of market allocations. To the contrary, there is a presumption that the government should try to divert resources to the growth-generating sector. Under Harold Wilson, Nickolas Kaldor had pushed for a selective employment tax to redirect resources towards manufacturing, which he believed exhibited greater returns to scale and more learning. (Kaldor, 1966)

We provide a parsimonious and tractable framework for examining these questions, a two-sector overlapping generations model with credit frictions, which enables us to contrast partial equilibrium short run effects with long run effects. Thus, lowering collateral requirements or interest rates might seem to enhance the ability of entrepreneurs to finance their investments; but if looser monetary policy or financial liberalization simultaneously results in an increase in land prices, investment may be diverted to real estate, and the net effect is to crowd out productive investments, lowering growth. We provide conditions under which the crowding out effect dominates. This will be so, for instance, when collateral requirements for real estate are lowered (not surprisingly), or even when the interest rate is lowered, when collateral requirements for real estate are lower than for manufacturing. Again, not surprisingly, we show that under such conditions, more rapid expansion of the money supply may not lead to higher growth, contrary to the suggestion of Tobin (1965). The reason is obvious: in his work, as with much of monetary economics, there are only two stores of value, money and capital; here, land is an alternative asset.

---

[1] Mian, Sufi, and Verner (2017) also found empirically that an increase in the household debt-to-GDP ratio predicts lower GDP growth.
[2] Chakraborty, Goldstein, and MacKinlay (2018) empirically studied the effect of housing prices on bank commercial lending and firm investment in the U.S. in the period between 1988 and 2006. They found housing price booms led to crowding-out of commercial investment due to decreased lending by banks to credit-constrained firms. They conclude that housing price booms have negative spillovers to the real economy.

Our model with credit frictions also provides a resolution to a longstanding conundrum: In standard balanced growth models, where land productivity increases with the growth of the economy, equilibrium only exists when the rate of interest exceeds the rate of growth, because otherwise, land prices would be infinite. But, as Blanchard (2019) and others have shown, in the past century of U.S data, the return to government debt has been lower than the growth rate of output.[3] Still, land prices are finite. In our model, the interest rate received by non-entrepreneurial households is less than the rate of growth, but that in turn is still less than the (leveraged rate of) return to land (which is equal to the leveraged return to capital).

While many of the assumptions employed in the model might seem stark—there is, for instance, full spillovers of productivity improvements from the productive sector to the real estate sector (as assumption employed to enable the easy construction of balanced growth), the real estate sector uses land only, and workers cannot invest in real estate—these are employed to simplify the analysis. Our key results are robust to weakening these assumptions. We provide a broader analysis of robustness in the on-line Appendices and show that the results can be generalized.

**Section 2. The basic model: endogenous growth with land**

**2.1 The basic environment**

We use a variant of the standard two-period overlapping generations model where there is a competitive economy with productive capital, land and labor. Time is discrete and extends from zero to infinity ($t = 0, 1, ...$). In each period agents arrive and live for two periods. Each agent is endowed with one unit of labor when young, and supplies it inelastically, receiving wage income, $w_t$. A fraction $\eta$ of the young are entrepreneurs who have investment opportunities. $R_{t+1}$ is the rental rate of capital at date $t + 1$, and $\delta$ is the depreciation rate of capital. The remaining fraction $1 - \eta$ are savers who don't have investment opportunities. To simplify the analysis, all agents consume only in the second period, i.e., $u_t^i = c_{t+1}^i \geq 0$, where $u_t^i$ is the utility of agent $i$, and $i$ is the index for each agent.[4]

---
[3] Implying in standard models that the economy is also dynamically inefficient, inconsistent with dynastic growth models.

There are two sectors in our model, a productive sector (in the simplified model of the text, the only sector that uses capital; more generally, the capital-intensive sector) and a real estate sector (which in the simplified version of our model, only uses land).

To illustrate the key ideas clearly, we first consider a small open economy where the interest rate $r$ is exogenously given. In section 3, we will develop a closed economy model with money where the interest rate is endogenously determined.

We focus on the case where only entrepreneurs participate in the real estate market. One interpretation of this assumption is that there is considerable land heterogeneity and workers are less informed about real estate (with high costs associated with obtaining the relevant information), making real estate unattractive for them.[5] In section 3 we will briefly discuss the case where both agents participate in the real estate market.

The budget constraint of agent $i$ is given by

(1) $\quad k_{t+1}^i + P_t x_t^i = w_t + b_t^i$ and $c_{t+1}^i = (R_{t+1} + 1 - \delta)k_{t+1}^i + \left(\frac{D_{t+1} + P_{t+1}}{P_t}\right) P_t x_t^i - (1 + r)b_t^i,$

where $k_{t+1}^i$ and $x_t^i$ are the entrepreneur's capital investment and land holdings at date $t$. $b_t^i$ is the amount of borrowing at date $t$ if $b_t^i > 0$, and lending if $b_t^i < 0$. $R_t^c \equiv R_{t+1} + 1 - \delta$ is the total return per unit of capital investment made at date $t$.[6] $P_t$ and $D_{t+1}$ are the price of land at

---
[4] In Appendix O1, we study a case in which each agent has log-utility over consumption in both young and old periods, respectively. We show that the results are unchanged.
[5] For instance, there are real estate ventures yielding a zero return and those yielding a positive return, and because non-entrepreneurial workers can't distinguish between them, they do not invest in real estate; the (risk-adjusted) expected return is just too low.

If there are a finite number of projects, the informed will select the good projects (on average), leaving a disproportionate share of bad projects for the uninformed, so that even if the expected returns of the informed is equal to or greater than other investment opportunities, that of the uninformed is less even than government bonds. This effect is strengthened if we had, more realistically, assumed all individuals are risk averse.

There are other reasons for limited participation in the real estate market. For instance, people have different attitudes toward risk. The real estate market is obviously more risky than the (real) return on government bonds, so very risk averse agents don't participate in the real estate market.
[6] That is, an individual who invests a dollar in capital today gets a return on that investment of $R_{t+1}$ and has an asset worth $1 - \delta$ at the end of the period.

date $t$ and land rents at date $t + 1$ in terms of consumption goods. $R_t^x$ is the unleveraged return to land,

(1a) $\quad R_t^x = \frac{D_{t+1}}{P_t} + \frac{P_{t+1}}{P_t},$

i.e., the return per dollar spent on land holdings without using borrowing between date $t$ and $t + 1$. It is a key endogenous variable, depending critically on $P_t$ and $P_{t+1}$.

Since workers don't have investment opportunities and don't participate in the real estate market, their behavior is simple, i.e., they lend all savings either to entrepreneurs and/or to abroad at the exogenously given an interest rate, $1 + r$.

We assume credit frictions. The entrepreneurs cannot borrow unless they have collateral. The borrowing constraint is

(2) $\quad (1 + r)b_t^i \leq \theta R_t^c k_{t+1}^i + \theta^x R_t^x P_t x_t^i,$

where the first and second terms of the right-hand side in (2) reflect that only a fraction $\theta \in [0,1]$ of the total returns from capital investment and only a fraction $\theta^x \in [0,1]$ of the total returns from land holdings can be used as collateral. (2) means that total debt repayment obligations cannot exceed the total value of the collateral, and there are no defaults.[7]

We now describe the production side of the economy. The productive sector exhibits standard Marshallian external increasing returns to scale in capital investment (Aoki 1970, 1971; Frankel 1962; Romer 1986.)[8] In the productive sector, competitive firms produce the

---

[7] See Stiglitz and Weiss (1986) and Hart and Moore (1994) and for a micro foundation of this type of debt contract.

[8] There is a large literature justifying the presence of these Marshallian increasing returns externalities, which we will not repeat; the existence of these effects is at the center of much of the endogenous growth literature. With scale economies, large scale enables high wages, which in turn has a feedback in allowing still larger scale, and with the right parameterization (the standard one, and the one we use here), this can generate steady endogenous growth. One big advantage of this approach is that it allows individual firms to face constant returns to scale—they ignore the aggregate scale benefit because their contribution to it is so small—while there are aggregate scale economies; thus, there can exist a competitive equilibrium, but of course, because of the externality, there is a presumption that it won't be efficient.

There is an alternative strand of research, growing out of Arrow's model of learning by doing (1962), where *improvements* in technology depend on gross investment today. That there are important differences should be clear: in the case of increasing returns, should something happen that would lead to less investment in the productive sector—say a decision by workers to consume more "housing" and less manufactured goods—then there could be a fall in productivity. By contrast, in the learning model, such a decision would simply slow the rate of increase of productivity. Productivity itself would not decline. Still, with appropriate parameterizations, one can get similar results out of the different formulations. (For examples of work in this tradition, see Stiglitz 1987, Stiglitz and Greenwald 2014, and Dasgupta and Stiglitz 1988).

final consumption goods by using capital and labor. To keep things simple, we assume that the production function of each firm $j$ is given by

(3a) $\quad y_{tj} = (k_{tj})^\alpha (\chi(K_t) l_{tj})^{1-\alpha}$,

where $k_{tj}$ and $l_{tj}$ are capital and labor inputs of firm $j$. $\chi(K_t)$ is labor productivity, with $\chi'(K_t) > 0$, where $K_t$ is aggregate capital stock at date $t$. When we aggregate over all firms, we get at the aggregate level returns to scale. We assume $\chi(K_t)$ takes on a particular functional form:

(4) $\quad \chi(K_t) = aK_t$.

This is a key (though conventional) simplifying assumption.

In the real estate sector, one unit of land produces $D_t$ units of consumption goods (in Appendix O2, we relax this assumption and examine a case where both labor and land are inputs for real estate). To ensure the existence of the balanced growth path, we focus on a case where there is a full spillover from the productive sector to the real estate sector, so that land productivity grows at the economy's growth rate,

(3b) $\quad D_t = \epsilon \chi(K_t) = \epsilon a K_t$,

where $\epsilon \geq 0$ is a parameter that captures the level of land productivity relative to labor productivity in the productive sector.[9] $\epsilon$ can also be interpreted as the parameter that captures the size of the spillover to the real estate sector from the productive sector; when $\epsilon$ is larger, with an increase in $K_t$, the productivity increase in the real estate sector is larger.

Apart from the introduction of land, this is the standard $AK$ model. Aggregate output $Y_t$ is then

(5) $\quad Y_t = (K_t)^\alpha (\chi(K_t) L_t)^{1-\alpha} + D_t X_t$,

---

[9] So long as the positive spillover effect on the real estate sector is a linear function of $K_t$, we can ensure the existence of a balanced growth path. This approach to achieving balanced growth is similar to that taken by Greenwald and Stiglitz (2006). More generally, to obtain balanced growth, we need to construct a model so that different sectors will at least eventually grow at the same rate. It is clear from our parametrizations that the rate of growth of productivity in the productive sector equals the rate of growth of productivity in real estate. In Appendix O3, we relax the assumption of a full spillover, i.e., imperfect spillovers, in which case unbalanced growth dynamics occurs. We show that even if we study unbalanced growth dynamics, the same results can be obtained.

The aggregate supply of land and labor force are fixed: $X_t = X$, and $L_t = L$. We normalize $L$ and $X$ at unity.[10]

## 2.2 The behaviour of individual firms

In the productive sector, the individual firm ignores its tiny influence on the aggregate capital stock and thus on the productivity of its own worker. Thus, each firm employs capital and labor up to the point where its private marginal product equals the rental rate of capital and the wage rate, respectively. Using (4), factor prices and the aggregate production function become (see Appendix A1 for derivations)

(6a) $R_t = \alpha A$,

(6b) $w_t = (1-\alpha)AK_t$,

(6c) $Y_t = AK_t + \epsilon a K_t = (1 + \epsilon(a)^\alpha)AK_t$,

where $A \equiv (a)^{1-\alpha}$. (6) holds at every momentary equilibrium, not just in steady state. The wage rate and aggregate output grow at the same rate of aggregate capital stock, so $\frac{w_{t+1}}{w_t} = \frac{K_{t+1}}{K_t} = \frac{Y_{t+1}}{Y_t}$ holds in steady state equilibrium, and the rental rate of capital is constant, with the increasing labor productivity as aggregate capital stock increases canceling out the effect of capital deepening and the diminishing returns associated with it.

## 2.3 The behavior of entrepreneurs and competitive equilibrium

Throughout our paper, we focus on the case where the borrowing constraint (2) binds for entrepreneurs in equilibrium at each date, so that

(7) $R^c \equiv R + 1 - \delta > 1 + r$

---

[10] As another interpretation, this production function corresponds to the limiting case of $\sigma \to \infty$ in the following CES production function. $Y_t = \left(\gamma_1((K_t)^\alpha(\chi(K_t)L_t)^{1-\alpha})^{\frac{\sigma-1}{\sigma}} + \gamma_2(\chi(K_t)X_t)^{\frac{\sigma-1}{\sigma}}\right)^{\frac{\sigma}{\sigma-1}}$, where $\sigma$ is the elasticity of substitution between $(K_t)^\alpha(\chi(K_t)L_t)^{1-\alpha}$ and $\chi(K_t)X_t$, and $\gamma_1$ and $\gamma_2$ are parameters and are set to be unity.

that is, the return on capital net of depreciation is greater than the safe rate of interest, implying that entrepreneurs would want to borrow as much as they could.[11] Using (6a) we have

(8) $\quad R^c = \alpha A + 1 - \delta,$

with $R^c$ being fixed and exogenous, depending just on the technological parameters $a$, $\alpha$, and $\delta$.

*Optimal portfolio allocations*

Entrepreneurs allocate their portfolio between capital and land, both using borrowing. Since they are perfect substitutes[12], in equilibrium, the leveraged rates of returns on each must be the same (see Appendix A2 for derivations):

(9) $\quad \underbrace{\dfrac{R^c(1-\theta)}{1-\dfrac{\theta R^c}{1+r}}}_{\substack{\text{leveraged rate of return}\\\text{on capital investment}}} = \underbrace{\dfrac{R_t^x(1-\theta^x)}{1-\dfrac{\theta^x R_t^x}{1+r}}}_{\substack{\text{leveraged rate of return}\\\text{on land speculation}}}$

(9) determines the equilibrium value of $R_t^x$, which we can solve for explicitly as

(10) $\quad R_t^x = \dfrac{\lambda}{1-\theta^x+\dfrac{\theta^x \lambda}{1+r}} \equiv R^{x*},$

where $\lambda \equiv \dfrac{R^c(1-\theta)}{1-\dfrac{\theta R^c}{1+r}}$ is the leveraged return on capital investment. $\lambda$ and $R^{x*}$ are fixed, but depend both on technology parameters and policy variables.

For the denominator of the expression for the leveraged rate of return on capital investment to be positive, $1 + r > \theta R^c = \theta(\alpha A + 1 - \delta)$. We require $\theta$ to be sufficiently small so that is the case, i.e.[13]

---

[11] Indeed, each small entrepreneur, believing that (s)he has no effect on $R^C$ or $r$, would want to borrow an infinite amount to buy an infinite amount of the capital good. This, of course, ignores risk. Realistically, for many entrepreneurs, risk aversion is sufficiently low (or negative) that the binding constraint is the borrowing constraint that we focus upon. See also discussions in section 5.

[12] The qualitative results would be similar if the two assets were imperfect substitutes, but the calculations would be much more complicated and less transparent.

[13] Entrepreneurs seek to arbitrage the low interest rate at which they can borrow and the high returns they get from their investment. Without borrowing constraints, or with too weak borrowing constraints, they would obviously borrow an infinite amount. Assumption 1 puts a bound on $\theta$ that prevents that from happening.

**Assumption 1.** $\theta < \frac{1+r}{R^c} = \frac{1+r}{\alpha A + 1 - \delta} < 1$

Recalling the definition of $R_t^x$, we obtain the first basic difference equation.

(11) $P_{t+1} = P_t R_t^x - D_{t+1} = P_t R_t^x - \epsilon \chi(K_{t+1}) = P_t R^{x*} - \epsilon a K_{t+1}$.

*The capital-investment function*

By substituting (2) into (1) and solving for $k_{t+1}^i$, we can derive the capital investment function of entrepreneurs when the borrowing constraint binds.

(12) $k_{t+1}^i = \underbrace{\frac{1}{1 - \frac{\theta R^c}{1+r}}}_{\text{leverage}} \left[ \underbrace{w_t}_{\text{income}} - \underbrace{\left(1 - \frac{\theta^x R^{x*}}{1+r}\right) P_t x_t^i}_{\substack{\text{total down-payment} \\ \text{in buying land}}} \right]$,

i.e., we can calculate an entrepreneur's capital holdings by taking his wage, subtracting what he has to pay to buy the land he holds (which depends on land leverage); and leveraging up that amount up (through borrowing). $1 - \frac{\theta^x R^{x*}}{1+r}$ is the down-payment of a unit of land purchase.[14]

*The competitive equilibrium*

The competitive equilibrium is defined as a set of prices $\{P_t, w_t, R_t, R_t^x\}_{t=0}^{\infty}$ and quantities $\{c_t^i, b_t^i, x_t^i, k_{t+1}^i, \int c_t^i di, \int b_t^i di, \int x_t^i di, \int k_t^i di, Y_t\}_{t=0}^{\infty}$, given an initial $K_0$, such that (i) each entrepreneur chooses land holdings, capital investment, and borrowing to maximize utility under the budget and the borrowing constraints, and (ii) each lender (non-entrepreneurial worker) lends all savings either to entrepreneurs and/or to abroad, and (iii) the market clearing conditions for land, capital, labor and goods are all satisfied. In the initial period $t = 0$, all land is held by the initial old entrepreneurs.

**2.4 Aggregate Dynamics**

---

When Assumption 1 and (9) hold, equilibrium leverage associated with land speculation $1 / \left(1 - \frac{\theta^x R_t^x}{1+r}\right)$ is also positive and finite.

[14] Using (10), we can show that it is always strictly positive, so long as $\theta^x < 1$.

We are now ready to derive aggregate dynamics. When we aggregate (12) across young entrepreneurs, we have

(13) $K_{t+1} = \frac{1}{1-\frac{\theta R^c}{1+r}}\left[\eta w_t - \left(1 - \frac{\theta^x R^{x*}}{1+r}\right)P_t\right] = \frac{1}{1-\frac{\theta R^c}{1+r}}\left[\eta(1-\alpha)AK_t - (1 - \frac{\theta^x R^{x*}}{1+r})P_t\right].$

Substituting (13) into (11) yields

(11a) $P_{t+1} = P_t R^{x*} - \frac{\epsilon a}{1-\frac{\theta R^c}{1+r}}\left[\eta(1-\alpha)AK_t - (1 - \frac{\theta^x R^{x*}}{1+r})P_t\right]$

$= P_t\left\{R^{x*} + \frac{\epsilon a\left(1-\frac{\theta^x R^{x*}}{1+r}\right)}{1-\frac{\theta R^c}{1+r}}\right\} - \frac{\epsilon a \eta(1-\alpha)A}{1-\frac{\theta R^c}{1+r}}K_t.$

The dynamics of this economy is characterized by (11a) and (13), which define $\{K_{t+1}, P_{t+1}\}$ as simple linear functions of $\{K_t, P_t\}$.

To solve the dynamics, we introduce a variable $\phi_t \equiv \frac{P_t}{AK_t}$, the ratio of the value of land relative to production in the productive sector. For shorthand, we refer to this as the relative size of land speculation. From the definition of $\phi_t$, we can derive the evolution of $\phi_t$.

(14a) $\phi_{t+1} = \left(\frac{P_{t+1}/P_t}{1+g_t}\right)\phi_t,$

where $1 + g_t = \frac{K_{t+1}}{K_t}$, i.e., $g_t$ is the growth rate of capital. The evolution of $\phi_t$ depends on the growth rate of output in the productive sector and that of land prices.

From the definition of $R_t^x$,

(1a') $\frac{P_{t+1}}{P_t} = R_t^x - \frac{D_{t+1}}{P_t} = R_t^x - \frac{\epsilon a K_{t+1}}{P_t} = R_t^x - \frac{\epsilon a}{A}\frac{AK_t}{P_t}\frac{K_{t+1}}{K_t} = R^{x*} - \epsilon(a)^\alpha \frac{1+g_t}{\phi_t}.$

From (13), we can calculate the growth rate as a function of $\phi_t$ and the various parameters of the problem:

(15a) $1 + g_t = \frac{A}{1-\frac{\theta R^c}{1+r}}\left[\eta(1-\alpha) - (1 - \frac{\theta^x R^{x*}}{1+r})\phi_t\right].$

Other things being constant, a rise in $\phi_t$ (land speculation) crowds out savings away from capital investment, reducing the growth rate of the economy. On the other hand, increases in $\theta$ or $\theta^x$, or a fall in $r$, increase leverage in capital investment and the down-payment of a unit of land purchase decreases, generating more capital investment. We say that capital investment is "crowded in." The question is what are the circumstances under which each effect dominates? Substituting (1a') into (14a), and using (15a) yields

(14b) $\phi_{t+1} = \frac{\phi_t R^{x*}}{1+g_t} - \epsilon(a)^\alpha = \phi_t R^{x*} / \left\{ \frac{A}{1-\frac{\theta R^c}{1+r}} \left[ \eta(1-\alpha) - (1 - \frac{\theta^x R^{x*}}{1+r})\phi_t \right] \right\} - \epsilon(a)^\alpha.$

$\phi_t$, given initial conditions, can be solved on its own, with the dynamics depending just on the technological parameters $A$ and $\epsilon$, on the share of the population that are entrepreneurs, $\eta$, and the market/policy parameters $\{r, \theta, \theta^x\}$. Much of the discussion below focuses on how changes in these parameters affect the economy's trajectory and long-run productivity and economic growth through both crowding-in and -out effects.

*Special case of unproductive land*

Consider the special case of $\epsilon = 0$, i.e., land is unproductive. Even so, land can still be used as collateral and as a store of value. It follows directly from (14b) that if there is a steady state[15], $1 + g^* = R^{x*}$; and (14b) becomes

(14c) $\phi_{t+1} = \phi_t R^{x*} / \left\{ \frac{1}{1-\frac{\theta R^c}{1+r}} \left[ \eta(1-\alpha)A - A(1 - \frac{\theta^x R^{x*}}{1+r})\phi_t \right] \right\}.$

The RHS is convex in $\phi_t$ and equals zero when $\phi_t = 0$, so there is a unique steady state with land having a positive price. The effects of changes in $r$ and collateral requirements on growth can easily be determined: they just depend on how those variables affect $R^{x*}$.

In this case, we can also solve easily explicitly for the steady-state equilibrium level of land speculation.

(16a) $\phi^* = \left\{ \eta(1-\alpha)A - R^{x*}(1 - \frac{\theta R^c}{1+r}) \right\} / A(1 - \frac{\theta^x R^{x*}}{1+r}).$

(There is another steady state with $\phi_t = 0$ for all $t$—effectively a "landless" economy. But as we will see below, such a "landless" equilibrium cannot exist with positive $\epsilon$.)

---

[15] The result follows more directly by observing that in a steady state $P_t$ must increase at the rate $1 + g^*$, and with $D = 0$, $P_t$ increases at the rate $R^{x*}$ from (1 a').

We impose an assumption concerning parameter values to ensure $\phi^* > 0$ even when $\epsilon = 0$.

**Assumption 2.** $R^{x*} < \frac{\eta(1-\alpha)A}{1-\frac{\theta R^C}{1+r}} \leftrightarrow \frac{(\alpha A+1-\delta)(1-\theta)}{1-\theta^x+\frac{\theta^x \lambda}{1+r}} < \eta(1-\alpha)A$

Under this assumption (which can easily be satisfied even when Assumption 1 holds), the slope of (14b) evaluated at $\phi_t = 0$ is strictly less than one.[16]

*General case and dynamics*

From the equilibrium condition (14b), if $\epsilon \neq 0$, then

(16b) $\phi^* = \left( Y_1 + \sqrt{Y_1^2 + 4\eta(1-\alpha)\left(\frac{A}{1-\frac{\theta R^C}{1+r}}\right)^2 \left(1 - \frac{\theta^x R^{x*}}{1+r}\right)\epsilon(a)^\alpha} \right) / \left( \frac{2A}{1-\frac{\theta R^C}{1+r}} \left(1 - \frac{\theta^x R^{x*}}{1+r}\right) \right),$

with $Y_1 = \frac{\eta(1-\alpha)A}{1-\frac{\theta R^C}{1+r}} - \frac{A}{1-\frac{\theta R^C}{1+r}}\left(1 - \frac{\theta^x R^{x*}}{1+r}\right)\epsilon(a)^\alpha - R^{x*}$. This gives $\phi^*$ as a function of the parameters of the model. As $\epsilon \to 0$, (16b)→(16a).

Since $\phi_{t+1}$ is a convex function of $\phi_t$, with the negative intercept when $\epsilon > 0$ and $\phi_{t+1} \to \infty$, as $\phi_t \to \bar{\phi} \equiv \eta(1-\alpha)/(1 - \frac{\theta^x R^{x*}}{1+r})$, there exists a unique and positive value of $\phi$ where $\phi_t = \phi_{t+1} \equiv \phi^* < \bar{\phi}$. Figure 1 illustrates this. Note that as $\phi_t \to \bar{\phi}$, $1 + g_t \to 0$ because the crowding-out effect becomes so large.

**Figure 1: Dynamics of $\phi_t$**

---

[16] If this condition is not satisfied, the only steady state entails $\phi^* = 0$.

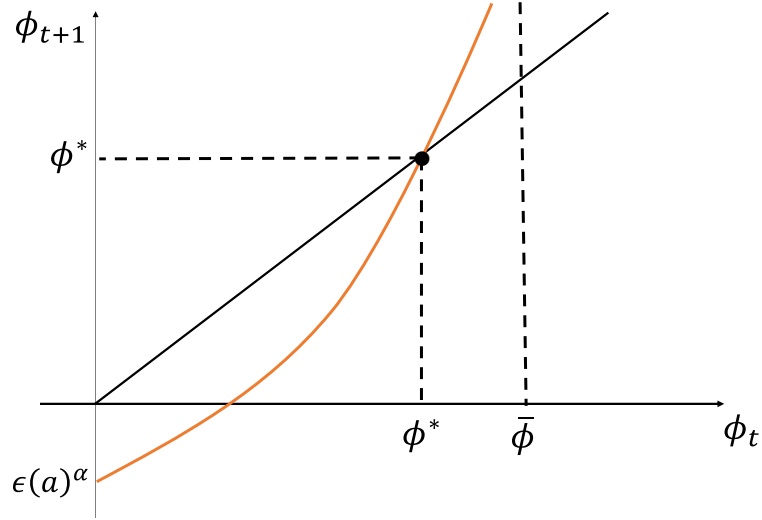

It is clear that unless $\phi_0$ is set at $\phi^*$, the economy does not converge. Given the initial value of $K_0$, there is a unique value of $P_0$ such that $\phi_0 = \phi^*$. $\phi_t$ remains at that value forever: there are no transitional dynamics for $\phi$.

Given this, $g_t$ is fixed:

(15b-1)  $1 + g_t = \dfrac{1}{1-\frac{\theta R^c}{1+r}}\left[\eta(1-\alpha) - (1 - \dfrac{\theta^x R^{x*}}{1+r})\phi^*\right] A \equiv 1 + g^*.$

Capital grows at a constant rate, and given the constancy of $\phi$, so does the price of land. For later reference, when the interest rate is constant, we note that the extent of spillover ($\epsilon$) affects $g^*$ only through its effects on the equilibrium level of land speculation, given by (16b).

Then we have the following Proposition.

**Proposition 1 Existence and Uniqueness of Rational Expectations Trajectory**

Under Assumption 1 and 2, for any $\epsilon \geq 0$, there exists a unique balanced growth path where the value of land relative to production in the productive sector $\phi^*$, the economic growth rate $1 + g^*$, and the growth rate of land prices $\dfrac{P_{t+1}}{P_t}$ are constant over time, with $\phi^*$ and $g^*$ given by (16) and (15b-1). The economy immediately converges to that path, through the setting of the initial price of land, $P_0 = AK_0\phi^*$. This is the unique rational expectations trajectory with positive land prices.

Moreover, using (14b), as $\epsilon \to 0$,

(15b-2) $1 + g^* \to R^{x*}$

This implies that in this limiting case, policy will affect the growth rate only through effects on $R^{x*}$.

We can then substitute (16b) into (15b-1) to derive $g^*$ as a function of all the parameters of the problem (see Appendix A3 for derivations):

(15b-3) $1 + g^* =$

$$\frac{R^{x*} + \frac{\eta(1-\alpha)A}{1-\frac{\theta R^C}{1+r}} + \epsilon(a)\alpha \frac{\left(1-\frac{\theta^x R^{x*}}{1+r}\right)A}{1-\frac{\theta R^C}{1+r}} - \sqrt{\left(\frac{\eta(1-\alpha)A}{1-\frac{\theta R^C}{1+r}} - R^{x*}\right)^2 + 2\epsilon(a)\alpha \frac{\left(1-\frac{\theta^x R^{x*}}{1+r}\right)A}{1-\frac{\theta R^C}{1+r}}\left(\frac{\eta(1-\alpha)A}{1-\frac{\theta R^C}{1+r}} + R^{x*}\right) + \left(\epsilon(a)\alpha \frac{\left(1-\frac{\theta^x R^{x*}}{1+r}\right)A}{1-\frac{\theta R^C}{1+r}}\right)^2}}{2}$$

.Since, even if there were no investment (investment is restricted to be non-negative[17], $1 + g^* > 1 - \delta$, (15b-3) implies that there is a critical value of $\epsilon \equiv \bar{\epsilon}$, above which $1 + g^* = 1 - \delta$, In the discussion below, we assume that $\epsilon$ is strictly less than $\bar{\epsilon}$.[18]

For completeness, we note that when $\epsilon = 0$, there is another steady state equilibrium with $\phi^* = 0$, where land is valued at zero, and with growth higher than in the case where land has a positive price; this equilibrium is stable; if $\phi_t < \phi^*$, the rational expetations trajectory converges to the landless economy. But, as noted already, such a "landless" equilibrium cannot exist with positive $\epsilon$.

*Comparative dynamics*

The variable of greatest interest is the growth rate. Using (15b-3), we can conduct various comparative dynamic exercises with changes in the market/policy parameters $\{r, \theta, \theta^x\}$. First, we study the case of small $\epsilon$, i.e., the real estate sector is sufficiently less productive compared to the capital-intensive sector. Straightforward differentiation in (15b-3) enables us to show the following Proposition.

**Proposition 2 Impact of changes in the collateral values on long-run economic growth**

---

[17] Equivalent to the assumption that capital goods can't be converted back to consumption goods.
[18] Later, we show that as $\epsilon$ increases, $\phi^*$ approaches $\bar{\phi}$, so $1 + g^*$ approaches zero. The constraint on $\epsilon$ is only satisfied if $\delta = 1$, i.e. capital fully depreciates one period of time.

(2-i) If $\epsilon$ is small, greater collateral values of land encourage land speculation with leverage, retarding long-run economic growth. That is, $\frac{d(1+g^*)}{d\theta^x} < 0$: The crowding out effect of land speculation overwhelms the crowding in effect.

(2-ii) $\frac{d(1+g^*)}{d\theta} > 0$.[19] With an increase in $\theta$, the crowding-in effects associated with greater collateral values of capital investment dominate the crowding-out effect, so the increase in $\theta$ leads to higher economic growth. This is true, even though the $\frac{Land\ values}{GDP}$ ratio rises in equilibrium, where $\frac{Land\ values}{GDP}$ ratio $\equiv \frac{P_t}{Y_t} = \frac{P_t}{(1+\epsilon(a)^\alpha)AK_t} = \frac{\phi^*}{(1+\epsilon(a)^\alpha)}$, as we shall shortly show.

The impact on long-run productivity and economic growth of an increase in liquidity (a relaxation of the collateral constraints) is markedly different depending on how it arises. If it is a result of relaxation in real estate financing, it can be productivity-and growth-retarding in the long run. On the other hand, if it rises mainly with the relaxation in capital investment financing, it will be productivity-and growth-enhancing.

**Proposition 3 Impact of changes in interest rates on long-run economic growth**

For small $\epsilon$, $sign \frac{d(1+g^*)}{dr} = sign(\theta^x - \theta)$.

This result follows directly from (15b-2) in Proposition 1, differentiating $R^{x*}$ with respect to $r$, given $\theta^x$ and $\theta$. If land is sufficiently unproductive, the long-run impact of low interest rates on economic growth depends on the relative size of $\theta^x$ and $\theta$. When $\theta^x > \theta$, the crowding-out effect caused by low interest rates (with land prices increasing) dominates the crowding-in effects; low interest rates encourage land speculation with leverage more than capital investment, reducing long-run productivity and economic growth. On the other hand, when $\theta^x < \theta$, just the opposite is true.

Intuitively, in economies where the collateral value of land is greater than that of capital investment, lower interest rates mean more funds flow into the real estate market, encouraging land speculation with leverage rather than increasing productive capital. Therefore, they can be harmful to long-run productivity growth and economic growth.

---

[19] Regarding the effect of a change in $\theta$, the sign of the derivative is always positive, regardless of the size of $\epsilon$.

The central message is clear: a loosening of monetary policy or of collateral requirements may so increase land speculation than resources are diverted from capital investment and growth may be impaired—consistent with the experiences in numerous countries and the studies cited in the introduction.

*Decomposing impacts of policy changes into partial and general equilibrium effects*

We obtain further insight into how growth is affected by changes in the interest rate or collateral requirements by decomposing the effects of changes into a direct (partial equilibrium) effect and a general equilibrium effect, where the latter represents the impact on the equilibrium level of speculation, through (15a):

$$(15a) \quad 1 + g^* = \underbrace{\frac{1}{1 - \frac{\theta R^c}{1+r}}}_{\text{PE effect 1}} \left( \eta(1-\alpha) - \underbrace{(1 - \frac{\theta^x R^{x*}}{1+r})}_{\text{PE effect 2}} \underbrace{\phi^*}_{\text{GE effect}} \right) A.$$

PE effect 1 represents the direct leverage effect, with investment and growth increasing with a rise in $\theta$ or with a decline in $r$. PE effect 2 represents the effect on capital investments through changes in the equilibrium down-payment in buying a unit of land. With increases in $\theta$[20] or $\theta^x$ or with a decrease in the interest rate, the equilibrium down-payment in buying land decreases, allowing the financing of more capital investments.

On the other hand, the GE effect represents the effect on capital investments as a result of changes in the relative size of land speculation. Straightforward differentiation of $\phi^*$ in (16b) yields

**Proposition 4 Impacts of changes in collateral requirements and interest rates on land speculation**

(4-i) $\frac{d\phi^*}{d\theta^x} > 0$ (4-ii) $\frac{d\phi^*}{dr} < 0$ (4-iii) $\frac{d\phi^*}{d\theta} > 0$ $(4-iv)$ $\frac{d\phi^*}{d\epsilon} > 0.$

In Figure 1, with increases in $\theta^x$ or $\theta$, or with a decline in $r$, $\phi_{t+1}$ is shifted down for any value of $\phi_t$, therefore raising $\phi^*$. Intuitively, making it easier to borrow, by lowering the interest rate or lowering collateral requirements, always increases the relative size of land speculation (reflected in $\phi^*$); and this is true even if the decrease in collateral requirements is limited to capital investment. When the relative size of land speculation increases, capital investments are reduced *cet. par*.

---

[20] If $\theta$ increases, a portfolio shift will occur from land speculation to capital investment. For the leveraged return on land and capital to remain equal, $R^{x*}$ must increase. As the unleveraged rate of return on a unit of land increases, it raises the collateral value of land, thereby reducing the equilibrium down-payment in buying a unit of land.

(4-iv) follows from (14b) an increase in $\epsilon$ shifts $\phi_{t+1}$ for any value of $\phi_t < \bar{\phi}$, thereby raising $\phi^*$. (See figure 1).

*Large $\epsilon$*

Using (15b-3), it is possible to ascertain the effects of changes in $\{r, \theta, \theta^x\}$ on $g^*$ for any value of $\epsilon < \bar{\epsilon}$. Since it is hard to provide a complete characterization analytically, we report results based on numerical calculations, at least for certain parameter values.[21] Figure 2-1 and 2-2 present numerical examples that show how the signs of $\frac{d(1+g^*)}{d\theta^x}$ and $\frac{d(1+g^*)}{dr}$ change as the spillover effect increases. In both cases, there exist critical values of $\epsilon$ above which the signs of the derivative get reversed, i.e., $\frac{d(1+g^*)}{d\theta^x} > 0$ and $\frac{d(1+g^*)}{dr} < 0$. In an economy where the spillover to the real estate sector from productivity increases is large and the productivity of the sector is high, a decline in $1 + r$ or an increase in $\theta^x$ could enhance long-run economic growth, i.e., the crowding-in effect could dominate the crowding-out effect. Intuitively, when $\epsilon$ is large, the equilibrium value of land speculation is large. With the increase in $\theta^x$, total down-payment in buying land decreases substantially, which enhances the crowding-in effect through PE effect 2, and the crowding-in effect can dominate the crowding-out effect. When the interest rate falls, we have another crowding-in effect through PE effect 1. Hence, the crowding-in effects are more likely to dominate the crowding out effect.

Figure 2-3 presents a numerical example that shows how the result in Figure 2-1 may change when the interest rate is lower. The range of $\epsilon$ for which $\frac{d(1+g^*)}{d\theta^x} < 0$ becomes wider. Indeed, $\frac{d(1+g^*)}{d\theta^x} < 0$ for all ranges in $\epsilon$ we show in the Figure. This implies that at a low interest rate environment, increasing land leverage is more likely to reduce productivity and economic growth.

**Figure 2-1**

---

[21] In our numerical calculations, we focus on the region where $1 + g^* \geq 1 - \delta$, which is a natural constraint.

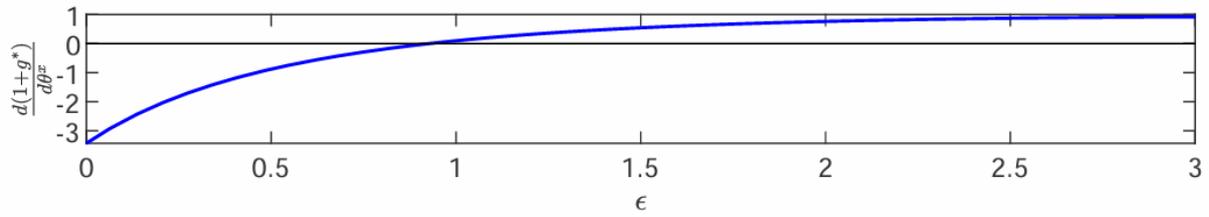

**Figure 2-2**

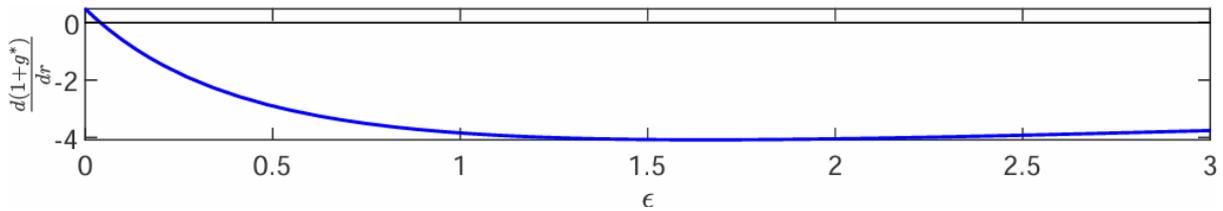

**Figure 2-3**

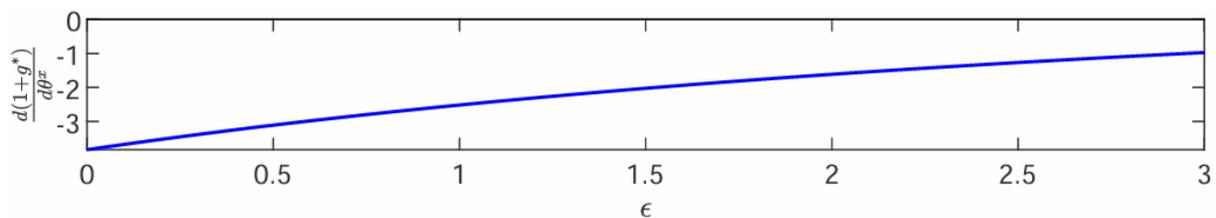

Note: Parameter values are set as $\theta^x = 0.6, \theta = 0.5, r = 0.55$ ($r = 0.44$ in Figure 2 − 3), $\eta = 0.4, \alpha = 0.33, a = 15.0, \delta = 0.2$.[22]

*Welfare analysis*

Given an initial capital stock $K_0$, and therefore the initial wage, the first generation's welfare, whether that of the worker or the entrepreneur, depends simply on $1 + r$ and the leveraged return, respectively, and increasing capital investment leverage increases the latter and leaves the former unchanged. Since that leads to a higher growth of $K$, and higher wages, *in every subsequent period,* workers are better off. But so are entrepreneurs. The older generation at time 0 are also better off: the policy change leads to a sudden increase in the value of land (given $K_0$, $\phi_0 = P_0/AK_0$ rises), and the entrepreneurs of that generation holding land receive a large capital gain. Hence, increasing capital investment leverage in this model is a Pareto improvement.[23] (There is a limit to this, of course, because of Assumption 1.)

---

[22] Since we employ the standard two-period overlapping generations model and consider the impact on long-run economic growth, it would be natural to interpret one period as a long period like twenty or thirty years. We leave a full-blown analysis of quantitative aspects as future work because the primary purpose of our paper is to obtain clear theoretical results.

But increasing land leverage has more ambiguous effects. In the 0$^{th}$ period, the price of land increases so the leveraged return to land remains equal to the (unchanged) leveraged return to capital. Neither workers nor entrepreneurs of the 0$^{th}$ generation are affected. If it results in lower growth (the conditions for which we have already identified), then wages *in every subsequent period* will be lower, so workers in every subsequent period are worse off. But the leveraged return on capital investment is unchanged, so that in spite of the higher land leverage, given the general equilibrium adjustments, entrepreneurs in every subsequent period are also worse off with the decrease in wages. *Increasing land leverage in these circumstances is almost Pareto inferior—everyone except the elderly at time 0 are worse off.* The elderly entrepreneurs, who hold the land, experience an unanticipated capital gain from a higher land price.[24]

**Section 3**. **Closed economy with money**

In this section, we extend the model to include fiat money and analyze a closed economy. Here, to ensure that there is a positive price for money, we make the further assumption that workers (a fraction $1 - \eta$ of the young generation) are endowed with $e_t^i = e_t$ units of consumption goods when young, while for entrepreneurs, $e_t^i = 0$.[25] We also assume (in order to have balanced growth) $e_t$ grows at the same rate as the growth rate of the economy, i.e., $e_t = e\chi(K_t)$. $T_t^i$ is a lump sum transfer payment from the government to old workers.[26] The analysis is much as before except now agents have an additional asset to hold.

---

[23] With output first period fixed, one might ask how can consumption of the old increase and investment by the young? The answer is that workers/savers lend less to foreigners.
[24] Similarly, a tax on total returns from land with its proceeds transferred to old workers induces a portfolio shift from land speculation to capital investment, which leads to reduced $\phi^*$, thereby increasing growth and the welfare of all generations except for the old generation who holds land when this policy change is introduced.
[25] We can, of course, allow $e_t^i > 0$ for entrepreneurs but since entrepreneurs don't hold fiat money in equilibrium, whether $e_t^i > 0$ for entrepreneurs is unimportant. Hence, we omit the case.

One can think of there being another productive asset (trees) whose output does not depend on labor or capital and therefore is unaffected by what goes on in the rest of the economy, except that its productivity increases in tandem with the rest of the economy.

(We need to introduce the assumption of worker endowment mainly for technical reasons, in establishing the existence of a monetary equilibrium, as the discussion below should make clear. It has no other substantive consequences.)

[26] We can also consider a case where government makes payments to old entrepreneurs. So long as $T_t^i$ cannot be pledged as collateral (e.g., social security payments are not allowed to be collateralized), our analysis applies as is; if it can be pledged as collateral, the analysis can be easily modified.

The budget constraint of entrepreneurial agent $i$ is given by

(1b) $k_{t+1}^i + P_t x_t^i + Q_t m_t^i = w_t + b_t^i$ and $c_{t+1}^i = R^C k_{t+1}^i + R_t^x P_t x_t^i - (1+r_t)b_t^i + Q_{t+1} m_t^i$,

where $m_t^i$ is the amount of money holdings of agent $i$ at date $t$ and where $Q_t$ is the price of fiat money in terms of consumption goods ($1/Q_t$ is the price level). The interest rate $1+r_t$ is now endogenously determined in equilibrium. We impose the short-sale constraint (agents can't hold a negative amount of money).

(3') $m_t^i \geq 0$,

Similarly, the budget constraint of workers can be written as

(1c) $Q_t m_t^i = w_t + e_t^i + b_t^i$ and $c_{t+1}^i = Q_{t+1} m_t^i - (1+r_t)b_t^i + T_t^i$,

Since there is no risk associated with workers holding fiat money or (by assumption) lending to entrepreneurs[27], in equilibrium, the rates of returns must be the same.

(17a) $1 + r_t = \frac{Q_{t+1}}{Q_t}$.

$\frac{Q_{t+1}}{Q_t}$ is the rate of return of a unit of fiat money. The (real) return to holding money, if it is positive, arises from deflation: money is more valuable tomorrow than today.[28]

Government issues fiat money according to

(17b) $M_{t+1} = (1+\mu)M_t$,

where $M_t$ is money supply at date $t$, and $\mu > 0$ is the growth rate of money supply. While in general, $\mu$ can vary period to period, in the steady state upon which we focus, $\mu$ is constant.

---

[27] If there is a finite support to the distribution of returns to entrepreneurial projects, the "zero" risk assumption can be translated into a maximum acceptable level of leverage.

[28] The equilibrium may be characterized by a negative real interest rate, in which case (17a) implies there is inflation. (Negative real interest rates are a sign of "oversaving," but in the absence of a store of value with a positive dividend, the equilibrium in a lifecycle model may entail such oversaving, as Diamond (1965) showed. Similar results hold if money is interest bearing or if there is a convenience yield associated with it. Then the equilibrium won't necessarily be characterized by deflation when r is positive.

Each entrepreneur chooses land holdings, capital investment, money holdings, and borrowing to maximize utility under the budget constraints and the borrowing and short-sale constraints, and each worker chooses the amount of money holdings and/or lending to entrepreneurs subject to the budget and short-sale constraints.

The government faces a budget constraint too; its borrowing plus money issue minus expenditures servicing the debt (interest payments plus repayments) plus taxes must equal expenditures (real expenditures plus transfer payments) at each date, and the present value of its indebtedness must go to zero. For simplicity, we assume the government does no borrowing, and thus must finance transfer payments out of money issuance, and the only transfer payments are made to old savers:

(18) $M_{t+1} - M_t = \mu M_t = \int T_t^i \, di$

In equilibrium all markets clear at all dates, including borrowing ($\int b_t^i di = 0$) and money markets ($\int m_t^i di = M_t$).

As before, we focus on the case where the borrowing and the short sale constraint bind,[29] in which case entrepreneurs do not hold fiat money because $R^c > 1 + r_t = \frac{Q_{t+1}}{Q_t}$.[30] Only workers hold fiat money in equilibrium. Entrepreneurs invest in real capital and land with maximum leverage.

The value of fiat money is determined so that savings each period (the income of young people, both entrepreneurs and workers) equals holdings of capital, land, and money:

(19) $Q_t M_t = w_t + (1-\eta)e_t - K_{t+1} - P_t$.

Using the definition of $\phi_t$, (19) can be written as

(20) $Q_t M_t = \left\{(1-\alpha) + (1-\eta)e(a)^\alpha - \frac{1}{1-\frac{\theta R^c}{1+r_t}}\left[\eta(1-\alpha) - (1 - \frac{\theta^x R_t^x}{1+r_t})\phi_t\right] - \phi_t\right\} AK_t,$

where now the interest rate is endogenous. So long as the term in the large bracket is positive, the price of fiat money will be positive. Intuitively, if the savings remaining after financing capital investment and land holdings are large enough, those idle savings flow toward money holdings. A sufficient condition for this is that $e$ be large enough, which we assume to be the case (see Appendix A4 for a formal argument.). In the analysis below, it will be clear that $g^*$ and $\phi^*$ do not depend on $e$; the price of money is determined residually, with no effect on the real values in the economy; the rate of expansion of the money supply, however, does matter.

---

[29] Using (22a-2), we can show that for any $\epsilon \geq 0$, the borrowing constraint binds at the steady state equilibrium i.e., $R^c > 1 + r^*$ if $\mu > 0$.

[30] In a sense, short selling money is no different from borrowing.

The dynamics of this economy, entailing not just the price of land and the capital stock, but now including $\{Q_t, M_t, r_t, T_t\}$ can be characterized by (14b), (15a), (17a), (17b), (18) and (20). First, we proceed with our analysis by focusing on steady states.

In steady state the real value of money grows according to

(21) $\quad 1 + r^* = \frac{Q_{t+1}}{Q_t} = \frac{1+g^*}{1+\mu}$.

As is standard in this class of models, whether inflation or deflation occurs in equilibrium depends on the relative size of $g^*$ and $\mu$. Hence, instead of $r$ being fixed, it is a simple function of $g^*$ and $\mu$. (15b-1), (16b), and (21) and give three equations in three unknowns, $r^*$, $g^*$, and $\phi^*$.

*Steady states: unproductive land*, i.e., $\epsilon = 0$

The solution when land is unproductive is particularly easy. The steady-state growth rate is still $1 + g^* = R^{x*} = \frac{\lambda}{1-\theta^x + \frac{\theta^x \lambda}{1+r^*}}$, where it will be recalled $\lambda \equiv \frac{R^c(1-\theta)}{1-\frac{\theta R^c}{1+r^*}}$. Using (21), we have

$(1 + r^*)(1 + \mu) = \frac{\lambda}{1-\theta^x + \frac{\theta^x \lambda}{1+r^*}}$, which, after much simplifying, can be solved for $r^*$, and hence $g^*$ and $\phi^*$:

(22a) $\quad 1 + r^* = \frac{R^c}{1-\theta^x}\left[\frac{1-\theta}{1+\mu} - (\theta^x - \theta)\right]$.

(22b) $\quad \phi^* = \frac{\eta(1-\alpha)}{1-\theta^x(1+\mu)} - \frac{R^c(1-\theta)}{A(1-\theta^x)}$.

(22c) $\quad 1 + g^* = \frac{R^c}{1-\theta^x}[1 - \theta - (\theta^x - \theta)(1 + \mu)]$.

i.e., steady state growth depends in a simple way on the leverage variables, the rate of growth of money, and the total return per unit of capital investment. For $\phi^*$ to be positive, we require that $1 + \mu$ lie between certain bounds. Assumptions 3 and 4 below guarantee that this is the case.

*General case*

The more general case is not solved so simple. By rearranging (15b-1), (16b), and (21), we can derive the quadratic equation for $\phi^*$ (see Appendix A5 for detail). We can show there that when $\epsilon > 0$, one of the solutions is positive and the other one is negative (when $\epsilon = 0$, the other one is zero). Solving $\phi^* > 0$ yields

(22b') $\quad \phi^* = \dfrac{-Y_2 + \sqrt{Y_2^2 + 4A[1-\theta^x(1+\mu)]\epsilon(a)^\alpha\left\{\eta(1-\alpha)A + A\theta^x(1+\mu)\epsilon(a)^\alpha + \frac{R^C(1-\theta)}{1-\theta^x}\theta^x(1+\mu)\right\}}}{2A[1-\theta^x(1+\mu)]}$,

where

$Y_2 = -\eta(1-\alpha)A - A\theta^x(1+\mu)\epsilon(a)^\alpha + A\epsilon(a)^\alpha[1-\theta^x(1+\mu)] + \dfrac{R^C(1-\theta)}{1-\theta^x}[1-\theta^x(1+\mu)]$.

(22a) and (22c) can also be written as

(22a') $\quad 1 + r^* = \dfrac{R^C}{1-\theta^x}\left[\dfrac{1-\theta}{(1+\mu)\left(1+\frac{\epsilon(a)^\alpha}{\phi^*}\right)} - (\theta^x - \theta)\right]$,

(22c') $\quad 1 + g^* = (1+r^*)(1+\mu) = \dfrac{R^C}{1-\theta^x}\left[\dfrac{1-\theta}{\left(1+\frac{\epsilon(a)^\alpha}{\phi^*}\right)} - (\theta^x - \theta)(1+\mu)\right]$.

To ensure that the equilibrium leverage associated with capital investment is finite, with $r$ endogenous, we replace Assumption 1 with

**Assumption 3.** $\quad 1 > \theta^x(1+\mu)\left(1 + \dfrac{\epsilon(a)^\alpha}{\phi^*}\right)$,

where $\phi^*$ is given by (22b').

We also impose parameter values to ensure $\phi^* > 0$, even when $\epsilon = 0$.

**Assumption 4.**  $\frac{\eta(1-\alpha)}{1-\theta^x(1+\mu)} > \frac{R^C(1-\theta)}{A(1-\theta^x)} = \frac{(\alpha A+1-\delta)(1-\theta)}{A(1-\theta^x)}$ [31]

Then we obtain the following Proposition.

**Proposition 5 Uniqueness of balanced growth path**

Under Assumption 3 and 4, for $\epsilon \geq 0$, there exists a unique balanced growth path where the relative size of land speculation $\phi^* > 0$, economic growth rate $1 + g^*$, the interest rate $1 + r^*$, which equals the rate of return on fiat money $\frac{Q_{t+1}}{Q_t}$, are constant over time.

Moreover, $g^*$, $\phi^*$, and $r^*$ are given by (22a'), (22b'), and (22c'), and they converge to the values given in (22a), (22b), and (22c), as $\epsilon \to 0$.

(22a'), (22b'), and (22c') allow us to conduct comparative statics with changes in the market/policy parameters. As before, we first analyze the case with $\epsilon$ sufficiently small.[32] Direct calculations yield the following two Propositions.

**Proposition 6 Impact of changes in the collateral values on long-run economic growth in a monetary economy (small $\epsilon$)**

(6-i) For $\epsilon$ sufficiently small, $\frac{d(1+g^*)}{d\theta^x} < 0$, with $\frac{d\phi^*}{d\theta^x} > 0$[33] and $\frac{d(1+r^*)}{d\theta^x} < 0$.

That is, with greater collateral values of land, land speculation increases, which produces a crowding-out effect. On the other hand, the interest rate declines[34], which increases equilibrium leverage with capital investment and decreases the down-payment on a unit of land, producing a crowding-in effect. Nonetheless, the crowding-out effect dominates the

---

[31] When $\epsilon = 0$, if Assumption 4 is not satisfied, the only steady state entails $\phi^*= 0$. It is easy to establish that there are values of $\theta^x$ which satisfy both Assumptions 3 and 4, at least for small $\epsilon$. Assumptions 3 and 4 can be expressed as providing minimum and maximum values of $1 + \mu$.

[32] As $\epsilon \to 0$, Assumption 3 $\to 1 > \theta^x(1 + \mu)$.

For equilibrium leverage to be finite at the steady state, the growth rate of money $\mu$ cannot be too high. If it is too high, the interest rate gets sufficiently low that equilibrium leverage will become infinite. The condition is obviously satisfied if $\theta^x$ and/or $\mu$ are small.

[33] From (22b), we have $\frac{d\phi^*}{d\theta^x} = \frac{\eta(1-\alpha)(1+\mu)}{[1-\theta^x(1+\mu)]^2} - \frac{R^C(1-\theta)}{A(1-\theta^x)^2}$. $\frac{d\phi^*}{d\theta^x} > 0$ if and only if $\frac{\eta(1-\alpha)(1+\mu)}{[1-\theta^x(1+\mu)]^2} > \frac{R^C(1-\theta)}{A(1-\theta^x)^2}$. This inequality condition can be written as $\frac{\eta(1-\alpha)}{[1-\theta^x(1+\mu)]} > \frac{(\alpha A+1-\delta)(1-\theta)}{A(1-\theta^x)} \frac{[1-\theta^x(1+\mu)]}{(1-\theta^x)(1+\mu)}$, which holds true under Assumption 4 because $\frac{[1-\theta^x(1+\mu)]}{(1-\theta^x)(1+\mu)} < 1$ if $\mu > 0$.

[34] From (21), with a decline in $g^*$, interest rates have to decline if there is to be a steady state.

crowding-in effect: the increase in land speculation is so great that productivity and economic growth are impaired even though the interest rate has decreased.

(6-ii) For $\epsilon$ sufficiently small, $\frac{d(1+g^*)}{d\theta} > 0$, with $\frac{d\phi^*}{d\theta} > 0$ and $\frac{d(1+r^*)}{d\theta} > 0$.

That is, an increase in the collateral value of capital investment increases leverage, which finances more capital investment, thereby increasing productivity and economic growth, even though there is an endogenous rise in the interest rate and in our measure of relative land speculation $\phi^*$.

**Proposition 7 Impact of monetary policy on long-run economic growth (small $\epsilon$)**

For $\epsilon$ sufficiently small, $sign \frac{d(1+g^*)}{d(1+\mu)} = sign(\theta - \theta^x)$, with $\frac{d\phi^*}{d(1+\mu)} > 0$ and $\frac{d(1+r^*)}{d(1+\mu)} < 0$.

That is, an increase in money growth rate reduces the interest rate and increases the relative size of land speculation. Whether long-run economic growth increases or decreases depends on the size of $\theta^x$ and $\theta$. When $\theta^x > \theta$, low interest rates induced by expansionary monetary policy encourage land speculation with leverage so much that productivity growth and economic growth decrease. On the other hand, when $\theta^x < \theta$, with the decline in the interest rate, more funds flow into capital investment rather than land speculation, thereby increasing productivity of the economy and enhancing long-run economic growth.

*Comparison with Tobin (1965)*

Proposition 7 is in sharp contrast with the conventional view of expansionary monetary policy in which there is a presumption that easier monetary and credit policies (e.g. a faster rate of expansion of the money supply) leads to more capital investment. (See, e.g., Tobin 1965) Tobin claimed that an increase in the money growth rate leads to higher inflation rates and reduced rates of return on fiat money, inducing a portfolio shift from fiat money to real capital, thereby leading to higher long run economic growth. This effect is called the "Tobin Effect".[35]

---

[35] There is, of course, a large literature identifying other effects on higher inflation, arguing that it is bad for economic growth. The simple growth model we present here is consonant with that investigated by Tobin, as we show below.

Tobin (and much of the related subsequent literature on money and growth) was correct that higher rates of money growth lead to higher rates of inflation; but he ignored the existence of land, and the shift away from money may (and often does) lead to a demand for more land rather than more real investment.

Our model can validate Tobin's claim in a landless economy, i.e., one where $\epsilon = 0$ and $P_t = 0$. From (15a), when $P_t = 0$, at the balanced growth path, the growth rate of the economy can be written as

$$(23) \quad 1 + g^* \equiv \frac{K_{t+1}}{K_t} = \frac{\eta(1-\alpha)A}{1 - \frac{\theta R^c}{1+r^*}}.$$

It is clear from (23) that a reduction in the interest rate relaxes the borrowing constraint, financing more capital investment and leading to higher economic growth. By substituting (23) into (21) and solving for $1 + r^*$, we have the standard result.

$$(24) \quad 1 + r^* = \frac{Q_{t+1}}{Q_t} = \frac{\eta(1-\alpha_1)A}{1+\mu} + \theta R^c.$$

That is, an increase in the money growth rate increases the inflation rates and lowers the long run interest rate. We summarize this result as the following Proposition.

**Proposition 8 Replication of the Tobin's claim in a model with credit frictions**

Consider a special case with $\epsilon = 0$ and $P_t = 0$, i.e., the landless economy. Then, for $\theta > 0$, we have $\frac{d(1+g^*)}{d(1+\mu)} > 0$, with $\frac{d(1+r^*)}{d(1+\mu)} < 0$.

Because decreased interest rates relax the borrowing constraint of entrepreneurs, more capital investment is financed and productivity and economic growth are higher.

But as the experience in the US in the beginning of the century illustrates, more expansionary monetary policy (lower interest rates) may lead to more real estate speculation—so much so that productive capital investment diminishes. This is what Proposition 7 shows. To ensure that the Tobin effect works, at the same time as monetary easing, financial regulations on the real estate sector (or taxes we discuss in conclusion) need also be introduced, i.e., $\theta > \theta^x$ so that more funds can flow into capital investment rather than land speculation.

*Dynamics*

The dynamics of this model are also similar to those of the small open economy. Looking at the difference equations for $\{Q_t, M_t, r_t, T_t\}$ [defined by (14b), (15a), (17a), (17b), (18) and (20)] it is clear that if the price of land at time 0 jumps to $P_0 = AK_0\phi^*$, and the price of money jumps to $Q_o = \left\{(1-\alpha) + (1-\eta)e(a)^\alpha - \frac{1}{1-\frac{\theta R^c}{1+r_*}}\left[\eta(1-\alpha) - \left(1 - \frac{\theta^x R^{x*}}{1+r^*}\right)\phi^*\right] - \phi^*\right\}\frac{AK_0}{M_0}$, then there is a steady state equilibrium where all the dynamical equations are satisfied at $1 + g^* \equiv \frac{K_{t+1}}{K_t}$, $1 + r^* = \frac{Q_{t+1}}{Q_t}$, and $\phi_t = \phi^*$ for all $t \geq 0$.

More generally, we show in Appendix A5 that the dynamics can be reduced to the two-dimensional dynamical system, with $\phi_{t+1} = \psi(\phi_t, 1+r_t)$ and $1 + r_{t+1} = \omega(\phi_t, 1+r_t)$, where $1 + r_t = \frac{Q_{t+1}}{Q_t}$ from (17a), and both $\phi_t$ and $1 + r_t$ are jump variables. We can show that there is a dynamic equilibrium path that is a "jump" to the steady state—just as there was in the small-open economy case. This rational expectations path always exists under our Assumptions 3 and 4.[36]

**Proposition 9**: **Existence of rational expectations path without transitional dynamics**

There exists a rational expectations path along which the economy achieves the balanced growth path immediately without transitional dynamics.

*Large $\epsilon$*

Next, we study the case with large $\epsilon$ when the interest rate is endogenously determined. (22c') allows us to ascertain the effects of changes in $\{\mu, \theta, \theta^x\}$ on $g^*$ globally. Since it is hard to obtain general analytic results, we solve the model numerically for global changes in $\epsilon$, for changes in $\theta$, $\theta^x$, or $\mu$ in a consistent manner with Assumption 3 and 4 and we focus on the region where $1 + g^* \geq 1 - \delta$.

Figure 3-1 and 3-2 present numerical examples that show $\frac{d(1+g^*)}{d\theta}$ and $\frac{d(1+g^*)}{d(1+\mu)}$ as $\epsilon$ gets larger. In both cases, we obtain results closely parallel to those of the previous section 2, that

---

[36] Though we show that, at least for certain parameter values, it is the only rational expectations convergent local trajectory, we have not shown that there might be other rational expectations trajectories but based on numerical calculations we did, there are no paths near the equilibrium path that converge.

is, $\frac{d(1+g^*)}{d\theta} > 0$ globally, and $\frac{d(1+g^*)}{d(1+\mu)} < 0$ if $\epsilon$ is sufficiently small, while for $\epsilon$ large enough, $\frac{d(1+g^*)}{d(1+\mu)} > 0$ even if $\theta^x > \theta$. The reasons for the reversal of sign are analogous to that presented earlier.

For changes in $\theta^x$, there is a marked difference from the case where the interest rate is constant. Figure 3-3 shows that $\frac{d(1+g^*)}{d\theta^x} < 0$ is more likely to occur even for $\epsilon$ large enough than in the case of fixed $r$. Intuitively, unlike the case where the interest rate is constant, when it is endogenously determined, with a higher value of $\epsilon$, the interest rate also decreases endogenously, which in turn encourages land speculation, thereby strengthening the crowding-out effect.[37] This implies that when the interest rate is endogenously determined, increasing land leverage tends to reduce long-run productivity and economic growth, even when the spillover effect is large.

**Figure 3-1**

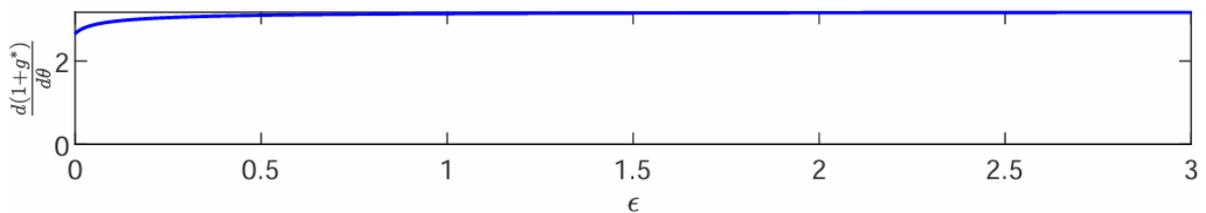

**Figure 3-2**

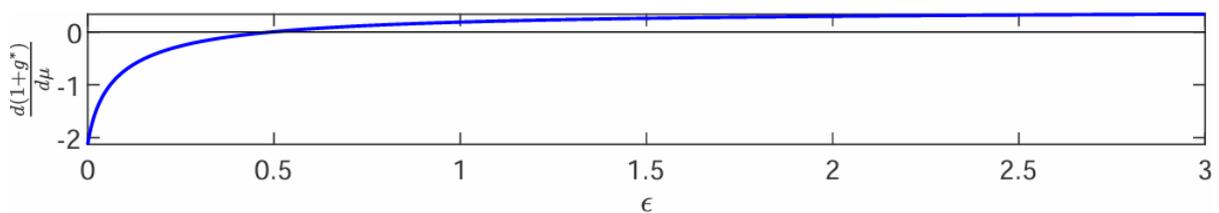

**Figure 3-3**

---

[37] This result is consistent with the result that when the interest rate is constant, the value of $\epsilon$ above which the sign of the derivative $\frac{d(1+g^*)}{d\theta^x}$ gets reversed increases when the interest rate is lower.

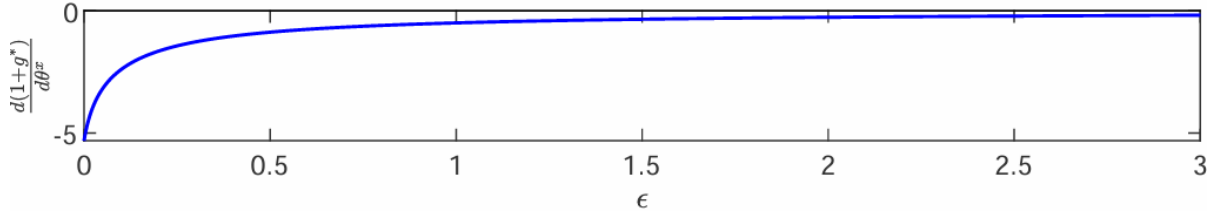

Note: Parameter values are set as $\theta^x = 0.6, \theta = 0.2, \mu = 0.5, \eta = 0.4, \alpha = 0.33, a = 15.0, \delta = 0.9$

**Section 5. Equilibrium with finite land prices, low interest rates, and high returns to capital**

In standard macroeconomic theory, if $r$ is very low, in particular less than the growth rate, there is a problem. There is dynamic inefficiency. Such an equilibrium can easily arise in life-cycle models without land. But if there exists an asset like land, with fixed returns no matter how small, as $r$ goes to zero, its market value becomes infinite, which of course cannot be an equilibrium. The only possible equilibria entail $r > g$.

But as we mentioned in the introduction, it has been the norm in the U.S economy that the return on government bonds have been less than the growth rate. The model we propose is at least theoretically consistent with this empirical observation.

At the balanced growth path, we know that $1 + r^* = \frac{1+g^*}{1+\mu}$ holds. It is obvious that $1 + g^* > 1 + r^*$ if $\mu > 0$. We also know from (14c) that $1 + g^* = R^{x*}/[1 + (\epsilon(a)^\alpha/\phi^*)]$, where $R^{x*} = \frac{\lambda}{1-\theta^x + \frac{\theta^x \lambda}{1+r^*}}$. So long as $\epsilon > 0$, $R^{x*} > 1 + g^*$. Moreover, $R^c > R^{x*}$ if $\theta^x > \theta$. Therefore, if $\theta^x > \theta$ and $\mu > 0$, $R^c > 1 + g^* > 1 + r^*$ holds at the balanced growth path. When $\epsilon = 0$, $1 + g^* = R^{x*}$ and the same argument applies. We summarize this result as the following Proposition.

**Proposition 10 Finite land prices with low interest rates**

If $\theta^x > \theta, \epsilon \geq 0$, and $\mu > 0$, $the\ return\ to\ capital > 1 + g^* > 1 + r^*$ holds at the balanced growth path, even though $\phi^*(P^*)$ is finite.[38]

---

[38] In Appendix O4, we study whether the equilibrium land price reflects fundamentals or contain bubbles.

To get this result, our assumption of limited participation in the real estate market plays an important role. Under this assumption, the discount rate in computing the discounted present value of land prices is $\frac{\lambda}{1-\theta^x+\frac{\theta^x\lambda}{1+r^*}}$, instead of $1+r^*$. If we consider the case where both entrepreneurs and savers participate in the real estate market, the no-arbitrage condition will be changed to $1+r_t = \frac{D_{t+1}}{P_t} + \frac{P_{t+1}}{P_t}$, in which case the discount rate will be $1+r^*$ on the balanced growth path.[39] If this is the case, land prices would become infinity when $1+r^* < 1+g^*$, as in standard models.

**Section 6. Discussions**[40]

**Discussion 1. Temporary output and asset price booms and their long-lasting negative effects on future generations**

So far, we have focused on the long-run effects of credit expansions or of low interest rate policy on productivity growth and economic growth. We can also examine how a temporary increase in productivity in the real estate sector has a long-run impact on the economy. Suppose that at the beginning of date $s$, there is an unexpected increase in productivity in the real estate sector, i.e., an increase in $\epsilon$.

Assume agents believe this shock to be permanent. In Figure 1, with this shock, $\phi_{t+1}$ is shifted down for any value of $\phi_t$, raising the equilibrium size of land speculation $\phi^*$. Land prices $P_s$ rise, given $K_s$. Since aggregate output and land prices at date $s$ simultaneously rise, this may appear to be an economic boom at the macroeconomic level. But at the sectoral level, the rise in $\phi^*$ crowds capital investment out in the productive sector, therefore reducing $K_{s+1}$, compared to when this shock did not occur. In our framework with endogenous growth, this decrease in capital investment, even if this productivity increase is temporary, will lower the level of capital accumulation permanently. In other words, even a temporary increase in productivity in the real estate sector produces crowding-out effects on productive capital, thereby leading to decreased wages and welfare of future generations, even if it generates temporary output and asset price booms. It should be noted that if the productivity increase is permanent, it generates a boom in output and land price for a while, but results in stagnant growth. Hence output and land price levels will therefore eventually be lower than they would have been had this shock not occurred.

---

[39] At the balanced growth path, since $1+r^* = \frac{1+g^*}{1+\mu}$ holds, we have $1+g^* > 1+r^*$ if $\mu > 0$. This implies that when we consider the case where both entrepreneurs and savers participate in the real estate market, for the existence of equilibrium, $\mu$ must be negative. Otherwise, land prices would become infinity.

[40] In Appendix A6, we study the effects of changes in $\{\mu, \theta, \theta^x\}$ on the Credit/GDP ratio. In O6, we also provide discussions concerning weakening the assumptions of limited participation in the credit market and credit frictions.

Assume, by contrast, that entrepreneurs realize that the real estate productivity shock is temporary. Then the no-arbitrage equation means that during the temporary real estate productivity boom, capital gains will be smaller. But that would mean that at $s + 1$, there would have to be a jump in the price of land. The only rational expectations trajectory entails the price of land rising at $s$, investment at $s$ falling (compared to the non-shock trajectory), and the price then at $s$ rising by less than it would have on the non-shock trajectory. Critically, on a rational expectations trajectory, even a temporary real estate boom has long-term negative hysteresis effects.

**Discussion 2. Related literature**

Our paper is related to the vast modern literatures on credit and financial frictions, land, monetary economics, OLG, and endogenous growth. It brings together in a parsimonious model key insights from each, noting how, in many instances, standard results are overturned once these more constrained models are extended. A key lesson is that we need to be careful in drawing implications for monetary policy in models without land; without endogenous growth; without a sensitivity to how monetary policy affects sectoral allocations (something that a model with a single sector obviously cannot do); and without credit frictions.

The modern macro-finance literature including seminal papers by Stiglitz and Weiss (1981), Bernanke (1983), and Diamond (1984) emphasizes the role of credit and the restrictions in the provision of credit to entrepreneurs. By putting credit at the center stage of macroeconomic analysis, the literature has deepened our understanding of how credit-driven macroeconomic fluctuations occur (see Greenwald, Stiglitz, and Weiss 1984; Stiglitz and Weiss 1986, 1992; Woodford 1988; Bernanke and Gertler 1989; and Greenwald and Stiglitz 1993 and Stiglitz and Greenwald 2003 for earlier work).

Based on these fundamental theoretical papers, Kiyotaki and Moore (1997) advanced the literature by incorporating asset prices explicitly (see also Bernanke, Gertler, and Gilchrist 1999). In their paper and many of subsequent papers,[41] land plays an important role as collateral. An increase in the value of land relaxes borrowing constraints of entrepreneurs who have productive investments, allowing them to make more of their productive investments, increasing the efficiency of macroeconomy. But it is assumed by model construction that the only thing productive entrepreneurs can do with borrowed funds is to engage in entrepreneurial activity with high returns. In contrast, in our model, entrepreneurs may use borrowed funds for land speculation as well as for productive investment. An increase in the collateral value of land produces a "crowding in" effect, as in their model, but through a quite different mechanism, i.e., in our model, a lower downpayment to buy real estate leaves more funds available for capital investment. More importantly, unlike their model, the increase in the collateral value of land produces a general equilibrium "crowding out" effect, as entrepreneurs' portfolios shift toward land speculation, rather than productive investments, which leads to increased land speculation, crowding out more productive capital investments. We provide the conditions under which the crowding out effect of land speculation outweighs the crowding in effect, suggesting a strong presumption that the former effect normally dominates, a result consistent with empirical results and common interpretations of many recent episodes of credit expansions (Verner 2019; Müller and Verner 2023).

Our paper is also related to the effect of land in overlapping generations models. There are two substantial differences of our paper from that literature. First, it is well known in the standard overlapping generations model with land that land holdings crowd out capital accumulation (Deaton and Laroque 2001; Mountford 2004). There is *only* a crowding out effect. Here, we introduce three more features to that standard model: there are two sectors, the real estate sector and the productive sector; and there is standard Marshallian external increasing returns to scale in capital investment, centered on the productive sector, that generates endogenous growth but whose benefits spillover into the rest of the economy; and there are "financial frictions"—limits in the extent to which entrepreneurs can borrow and workers can engage in real estate speculation. In this setting, more realistic than that of the standard OLG models, we show that land speculation may not only crowd out productive investment, but also lead to lower *rates* of growth. Credit expansions caused by increases in the collateral values of land or by a low interest rate policy are more likely to have negative growth effects than those induced by relaxing the collateral requirement on capital. Second, the welfare implications of the existence of land in our paper are markedly different from those in standard overlapping generations models with land. In the standard model with exogenous growth (including no growth), the existence of land eliminates inefficient equilibria (resolves an over-savings problem raised by Diamond 1965), increasing welfare (McCallum 1987). In contrast, in our model with endogenous growth, the very existence of land may reduce welfare by lowering rates of growth. Most importantly, monetary policies and financial de-regulations that expand credit to the real estate sector can be harmful to long-run growth and welfare.

---

[41] More recent macro-finance papers include Brunnermeier and Sannikov (2014) who analyse global dynamics in a macroeconomic model with financial sector, and Hirano and Yanagawa (2017) who study growth effects of asset price bubbles and the effects of their collapse on recovery, and Bolton, Santos, and Scheinkman (2021) who show that easy financial conditions lead to financial fragility, and Allen, Barlevy, and Gale (2022) who show that by introducing costly default exogenously, borrowers undertake excessively risky investments during asset booms and misallocation of resources occurs. A comprehensive literature review is too large for this paper; for a more complete discussion, see the 2022 Nobel Memorial Prize Economics https://www.nobelprize.org/uploads/2022/10/advanced-economicsciencesprize2022-2.pdf

In terms of a crowding-out effect and its impact on welfare, our paper is related to Ball and Mankiw (2023), who develop neoclassical growth models (exogenous growth) with market power. They show that market power can push the real interest rate below the economy's growth rate, in which case a Ponzi scheme of government debt may reduce steady-state welfare by crowding out capital. Our paper emphasizes crowding-out effects of land speculation with credit expansions on long-run and economic growth, while the relation of the return to capital > the rate of output growth > the safe rate of interest holds.

**Conclusion**

This paper begins with the hypothesis that some sectors of the economy generate more learning (more economies of scale, more likely to enhance endogenous growth) than others, and that the productivity benefits of that sector spillover to others.[42] Because of these within and cross-sector externalities, market allocations are not likely to be efficient. There is room for government intervention. For instance, here, given that the source of productivity growth is the "productive" (manufacturing) sector, it seems natural to expand that sector at the expense of the real estate sector. That, effectively, is what credit expansion does when there are weaker collateral requirements there, but it does just the opposite when there are tighter collateral requirements.

---

[42] In our review of the literature, we omitted a discussion of the vast literature on endogenous growth, either that based on learning-by-doing or that (as here) related to agglomeration economies. Here, we note the smaller literature on cross-sector spillovers and differential learning. See, in particular, Greenwald and Stiglitz (2006) and Stiglitz and Greenwald (2014) and the works cited there.

Intervention can, of course, take many other forms. For instance, government could impose a tax on the real estate sector, so that the return to holding land is reduced to $(1 - \tau)D$, with the revenues used to subsidize the productive sector, increasing after tax "output" to $(1 + \hat{\tau})Y$, where $\tau$ and $\hat{\tau}$ are chosen to balance the budget, effectively increasing $A$. Direct calculations from (16b) show that for $\epsilon$ small, an increase in $\tau$ leads a reduction in equilibrium land speculation $\phi^*$, and the direct effect of that is, of course, to increase $g^*$. Again, for small $\epsilon$, the effective increase in $A$ as a result of the subsidy to manufacturing also increases $g^*$.

There are two central messages of this paper. First, it is not so much aggregate credit expansion that matters for long-run productivity and economic growth but sectoral credit expansions. Credit expansions associated mainly with relaxation of real estate financing (of capital investment financing) will be productivity-and growth-reducing (enhancing) in the long run. An extension of our model can be used to derive a theoretical rationale for the role of ex-ante tighter financial regulations during credit expansions (see Lorenzoni 2008; Jeanne and Korinek 2019).

Second, and relatedly, lower interest rates may (as happened in the run-up to the 2008 financial crisis) induce more real estate speculation than productive capital investment; indeed the latter may be reduced, and with it economic growth—just the opposite of what we intended and predicted by many of the standard models which ignored the presence of land as an alternative (unproductive) store of value. Hence, to ensure enhanced growth, looser monetary policy has to be accompanied by financial regulations that ensure that funds flow into the productive sectors of the economy.

**Appendix**

**A1. Derivations of (6), (7), and (8)**

The production function of each firm $j$ is given by (3). Each firm maximizes its profit $\pi_{tj}$ by choosing $k_{tj}$ and $n_{tj}$, respectively. That is, the profit is given by

(A1-1)  $\pi_{tj} = y_{tj} - w_t n_{tj} - R_t k_{tj}$

Each firm chooses $k_{tj}$ and $n_{tj}$, taking $\chi(K_t)$ as given.

From the first-order necessary conditions, we can derive the demand equations for capital, labor, and land of each firm $j$, respectively, which we write as:

(A1-2) $\quad R_t = \alpha(k_{tj})^{\alpha-1}(\chi(K_t)l_{tj})^{1-\alpha}$

(A1-3) $\quad w_t = (1-\alpha)(k_{tj})^{\alpha}(\chi(K_t)l_{tj})^{-\alpha}\chi(K_t)$

Due to homogeneity of degree one, (A1-2), (A1-3), and (4) become

(A1-4) $\quad R_t = \alpha(K_t)^{\alpha-1}(\chi(K_t)L_t)^{1-\alpha}$

(A1-5) $\quad w_t = (1-\alpha)(K_t)^{\alpha}(\chi(K_t)L_t)^{-\alpha}\chi(K_t)$

(A1-6) $\quad Y_t = (K_t)^{\alpha}(\chi(K_t)L_t)^{1-\alpha}$

Substituting (4) into (A1-4), (A1-5), and (A1-6) respectively, yields (6), (7) and (8).

## A2. Deriving optimal portfolio selection

We construct the Lagrangian function,

(A2-1) $\quad \mathcal{L}_t^i = R^c k_{t+1}^i + R_t^x P_t x_t^i - (1+r)(k_{t+1}^i + P_t x_t^i - w_t)$

$\qquad\qquad + \Lambda_t^i[\theta R^c k_{t+1}^i + \theta^x R_t^x P_t x_t^i - (1+r)(k_{t+1}^i + P_t x_t^i - w_t)]$

where $\Lambda_t^i \geq 0$ is the Lagrangian multiplier associated with the borrowing constraint.

The first-order necessary conditions are:

(A2-2) $\quad \frac{\partial \mathcal{L}_t^i}{\partial k_{t+1}^i} = R^c - (1+r) + \Lambda_t^i[\theta R^c - (1+r)] = 0$

(A2-3) $\quad \frac{\partial \mathcal{L}_t^i}{\partial P_t x_t^i} = R_t^x - (1+r) + \Lambda_t^i[\theta^x R_t^x - (1+r)] = 0$

Complementary slackness conditions:

(A2-4) $\quad \Lambda_t^i[\theta R^c k_{t+1}^i + \theta^x R_t^x P_t x_t^i - (1+r)(k_{t+1}^i + P_t x_t^i - w_t)] = 0$ and $\Lambda_t^i \geq 0$

By rearranging (A2-2), we have

(A2-5) $\Lambda_t^i = \frac{R^c - (1+r)}{1+r-\theta R^c}$

Similarly, from (A2-3), we have

(A2-6) $\Lambda_t^i = \frac{R_t^x - (1+r)}{1+r-\theta^x R_t^x}$

It follows that if and only if $R^c > (1+r)$ and $R_t^x > (1+r)$, is $\Lambda_t^i > 0$, i.e., the borrowing constraint binds, provided that $1+r > \theta R^c$ and $1+r > \theta^x R_t^x$, which is Assumption 1.

From (A2-5) and (A2-6),

(A2-7) $\frac{R^c - (1+r)}{1+r-\theta R^c} = \frac{R_t^x - (1+r)}{1+r-\theta^x R_t^x}$,

which can be written as

(A2-8) $\frac{R^c(1-\theta) - (1+r) + \theta R^c}{1+r-\theta R^c} = \frac{R_t^x(1-\theta^x) - (1+r) + \theta^x R_t^x}{1+r-\theta^x R_t^x}$,

which is equivalent to (9),

## A3. Derivations for (15b-1) (growth rate in exogenous interest rate model)

From (14b) and (15b-1), we can derive the quadratic equation regarding $1 + g^*$.

(A3-1) $(1+g^*)^2 + \left(-R^{x*} - \frac{\eta(1-\alpha)A}{1-\frac{\theta R^c}{1+r}} - \epsilon(a)^\alpha \frac{\left(1-\frac{\theta^x R^{x*}}{1+r}\right)A}{1-\frac{\theta R^c}{1+r}}\right)(1+g^*) + \frac{\eta(1-\alpha)AR^{x*}}{1-\frac{\theta R^c}{1+r}} = 0$.

The solution is given by

(A3-2) $1 + g^* = \frac{-\zeta \pm \sqrt{\zeta^2 - 4\frac{\eta(1-\alpha)AR^{x*}}{1-\frac{\theta R^c}{1+r}}}}{2}$,

where $\zeta \equiv -R^{x*} - \frac{\eta(1-\alpha)A}{1-\frac{\theta R^c}{1+r}} - \epsilon(a)^\alpha \frac{\left(1-\frac{\theta^x R^{x*}}{1+r}\right)A}{1-\frac{\theta R^c}{1+r}}$.

When $\epsilon > 0$, one solution for $\phi^*$ is positive and given by (16b), while the other is negative. The only meaningful solution entails $\phi^* > 0$, obtaining (15b-3).

Note that when $\epsilon = 0$, (15b-2) can be written as

$$1 + g^* = \frac{R^{x*} + \frac{\eta(1-\alpha)A}{1-\frac{\theta R^c}{1+r}} \pm \sqrt{\left(R^{x*} + \frac{\eta(1-\alpha)A}{1-\frac{\theta R^c}{1+r}}\right)^2 - 4\frac{\eta(1-\alpha)AR^{x*}}{1-\frac{\theta R^c}{1+r}}}}{2} = \frac{R^{x*} + \frac{\eta(1-\alpha)A}{1-\frac{\theta R^c}{1+r}} \pm \sqrt{\left(\frac{\eta(1-\alpha)A}{1-\frac{\theta R^c}{1+r}} - R^{x*}\right)^2}}{2},$$

where $R^{x*} - \frac{\eta(1-\alpha)A}{1-\frac{\theta R^c}{1+r}} < 0$ under Assumption 2.

Therefore, we have two solutions,

$$1 + g^* = R^{x*} \text{ or } 1 + g^* = \frac{\eta(1-\alpha)A}{1-\frac{\theta R^c}{1+r}}.$$

$1 + g^* = R^{x*}$ is the solution corresponding to $\phi^* > 0$, while $1 + g^* = \frac{\eta(1-\alpha)A}{1-\frac{\theta R^c}{1+r}}$ is the solution corresponding to $\phi^* = 0$.

### A4. The existence of a unique balanced growth path in closed economy model

At the balanced growth path, we know that

(21) $\quad 1 + r^* = \frac{1+g^*}{1+\mu}.$

(14c) $\quad 1 + g^* = \frac{\lambda}{1-\theta^x + \frac{\theta^x \lambda}{1+r^*}} / (1 + \epsilon(a)^\alpha/\phi^*).$

where, it will be recalled, $\lambda \equiv \frac{R^c(1-\theta)}{1-\frac{\theta R^c}{1+r^*}}$. It follows from straightforward substitution that

(A4-1) $\quad 1 + r^* = \frac{R^C}{1-\theta^x}\left[\frac{1-\theta}{(1+\mu)(1+\epsilon(a)^\alpha/\phi^*)} - (\theta^x - \theta)\right]$

Using (A4-1), we have

(A4-2) $\quad 1 - \theta^x + \frac{\theta^x \lambda}{1+r^*} = \frac{1-\theta^x}{1-\theta^x(1+\mu)(1+\epsilon(a)^\alpha/\phi^*)}$

Substituting (A4-2) into (14c) yields the quadratic equations regarding $\phi^*$.

(A4-3) $A[1 - \theta^x(1+\mu)](\phi^*)^2 + \{-\eta(1-\alpha)A - A\theta^x(1+\mu)\epsilon(a)^\alpha + A\epsilon(a)^\alpha[1 - \theta^x(1+\mu)] + \frac{R^c(1-\theta)}{1-\theta^x}[1-\theta^x(1+\mu)]\}\phi^* - \epsilon(a)^\alpha\{\eta(1-\alpha)A + A\theta^x(1+\mu)\epsilon(a)^\alpha + \frac{R^c(1-\theta)}{1-\theta^x}[1-\theta^x(1+\mu)]\} = 0$.

Since the left-hand side of (A4-3) is a convex function of $\phi^*$, with the negative value at $\phi^* = 0$, it is clear from (A4-3) that if $\epsilon > 0$, one of the solutions is positive and the other one is negative. When $\epsilon = 0$, the other solution is zero. Hence, for $\epsilon \geq 0$, there exists a unique balanced growth path in which $\phi^* > 0$ is constant. It is obvious from (A4-1) and (14c) that when $\phi^* > 0$ is constant, $1 + r^*$ and $1 + g^*$ are constant and are uniquely determined.

We can also verify that for $e$ large enough, the price of fiat money be positive. From (A4-1) and (A4-3), we know that $\phi^*$ and $1 + r^*$ are independent of $e$. Substituting (15b) yields (20). For $e$ large enough, the term in the bracket is positive, i.e., the demand for fiat money is positive.

## A5. Dynamic stability of closed economy with money

As noted in the main text, the dynamics of this economy can be characterized by (14b), (15a), (17a), (17b), (18) and (20).

Using (10), (14b) can be written as

(A5-1) $\phi_{t+1} = \left( \frac{\frac{\lambda_t}{1-\theta^x + \frac{\theta^x \lambda_t}{1+r_t}}}{\frac{1}{1-\frac{\theta R^c}{1+r_t}}\left[\eta(1-\alpha) - (\frac{1-\theta^x}{1-\theta^x + \frac{\theta^x \lambda_t}{1+r_t}})\phi_t\right]A} \right) \phi_t - \epsilon(a)^\alpha = \psi(\phi_t, 1+r_t)$.

where $\lambda_t \equiv \frac{R^c(1-\theta)}{1-\frac{\theta R^c}{1+r_t}}$.

Using (15a), (17), and (20), (17a) can be written as

(A5-2) $1 + r_t = \frac{Q_{t+1}}{Q_t} =$

$$\frac{1}{1+\mu} \left[ \frac{\left\{(1-\alpha)+(1-\eta)e(a)^\alpha - \frac{1}{1-\frac{\theta R^c}{1+r_{t+1}}}\left[\eta(1-\alpha)-\left(\frac{1-\theta^x}{1-\theta^x+\frac{\theta^x \lambda_{t+1}}{1+r_{t+1}}}\right)\phi_{t+1}\right]-\phi_{t+1}\right\}}{\left\{(1-\alpha)+(1-\eta)e(a)^\alpha - \frac{1}{1-\frac{\theta R^c}{1+r_t}}\left[\eta(1-\alpha)-\left(\frac{1-\theta^x}{1-\theta^x+\frac{\theta^x \lambda_t}{1+r_t}}\right)\phi_t\right]-\phi_t\right\}} \right] \left\{ \frac{1}{1-\frac{\theta R^c}{1+r_t}}\left[\eta(1-\alpha) - \left(\frac{1-\theta^x}{1-\theta^x+\frac{\theta^x \lambda_t}{1+r_t}}\right)\phi_t\right] A \right\},$$

giving $r_{t+1}$ as a function of $\{r_t, \phi_t, \phi_{t+1}\}$. Using (A5-1), we have

(A5-3) $1 + r_{t+1} = \omega(\phi_t, 1 + r_t).$

Hence, the dynamics of the monetary economy can be characterized by (A5-1) and (A5-3), so it is reduced to the two-dimensional dynamical system, where both $\phi_t$ and $1 + r_t$ are jump variables.

We know from the analysis in the Appendix A4 that there exists a unique balanced growth path (steady state). Hence, there exists a rational expectation path along which the economy instantly achieves the balanced growth path. We have not been able so far to ascertain whether there might exist another rational expectations path.

**A6. Credit/GDP ratio**

When the borrowing constraint binds, the Credit/GDP ratio on the balanced growth path can be written as

(A6-1) $\frac{B_t}{Y_t} = \frac{\theta R^c}{1+r^*} \frac{K_{t+1}}{Y_t} + \frac{\theta^x R^x}{1+r^*} \frac{P_t}{Y_t}.$

Since $1 + r^* = \frac{1+g^*}{1+\mu}$ and $\frac{\theta^x R^x}{1+r^*} = \theta^x(1+\frac{\epsilon(a)^\alpha}{\phi^*})(1+\mu)$ on the balanced growth path, (A6-1) can be written as

(A6-2) $\frac{B_t}{Y_t} = \frac{\theta R^c(1+\mu)}{(1+\epsilon(a)^\alpha)A} + \frac{\theta^x(\phi^*+\epsilon(a)^\alpha)(1+\mu)}{1+\epsilon(a)^\alpha}.$

We know from Proposition 6 that $\phi^*$ is increasing functions of $\theta$ or $\theta^x$ or $1 + \mu$ for small $\epsilon$. Hence, we obtain the following Proposition.

**Proposition 11** For $\epsilon$ sufficiently small, the Credit/GDP ratio defined as $\frac{B_t}{Y_t}$ rises as $\theta$ or $\theta^x$ or $1 + \mu$ increases.

What drives our result is that faster expansion of the money supply lowers interest rates, effectively loosening the borrowing constraint.

This result, together with Proposition 6 and 7, has empirical implications. With those increases in the collateral values or a reduction in the interest rate caused by an increase in the money growth rate, Credit/GDP ratios rise but their impact on long-run productivity growth and economic growth can be markedly different depending on the source of the credit expansion. If the ratios increase mainly with the relaxation in real estate financing (in capital investment financing), productivity and economic growth slow down (speed up) in the long run. Sectoral credit allocation plays a key role for productivity and economic growth.

**Online Appendix**

In the main text, we presented a bare-bones model integrating credit frictions and endogenous growth. For tractability, we have made a number of simplified assumptions concerning preferences and technology. For instance, there is full spillovers of productivity improvements from the productive sector to the real estate sector, the real estate sector uses land only, agents have linear utility over consumption only when old, and workers cannot invest in real estate. In this on-line appendices, we show that the results can be generalized.

**O1. Robustness: More General Utility Function**

In this Appendix, we will consider a case with log utility in which each agent consumes in both young and old periods, respectively. The utility function is given by

(O1-1)   $u_t^i = \log(c_t^i) + \beta \log(c_{t+1}^i),$

where $c_t^i$ is consumption of a young person.

The budget constraints of agent $i$ in the young and old periods are

(O1-2a) $c_t^i + k_{t+1}^i + P_t x_t^i = w_t + b_t^i$

and

(O1-2b) $c_{t+1}^i = (R_{t+1} + 1 - \delta)k_{t+1}^i + \left(\frac{D_{t+1}+P_{t+1}}{P_t}\right) P_t x_t^i - (1+r)b_t^i.$

Each young person maximizes his/her utility (O1-1) subjects to (O1-2a) and (O1-2b). When we construct the Lagrangian function and substituting (O1-2) into (O1-1), we have

(O1-3) $\mathcal{L}_t^i = \log(w_t + b_t^i - k_{t+1}^i - P_t x_t^i) + \beta \log(k_{t+1}^i + R_t^x P_t x_t^i - (1+r)(k_{t+1}^i + P_t x_t^i - w_t)) + \Lambda_t^i[\theta R^c k_{t+1}^i + \theta^x R_t^x P_t x_t^i - (1+r)(k_{t+1}^i + P_t x_t^i - w_t)],$

where $\Lambda_t^i \geq 0$ is the Lagrangian multiplier imposed on the borrowing constraint.

The first-order necessary conditions:

(O1-4) $\frac{\partial L_t^i}{\partial k_{t+1}^i} = -\frac{1}{c_t^i} + \beta \frac{R^c}{c_{t+1}^i} + \Lambda_t^i \theta R^c = 0.$

(O1-5) $\frac{\partial L_t^i}{\partial P_t x_t^i} = -\frac{1}{c_t^i} + \beta \frac{R_t^x}{c_{t+1}^i} + \Lambda_t^i \theta^x R_t^x = 0.$

(O1-6) $\frac{\partial L_t^i}{\partial b_t^i} = \frac{1}{c_t^i} - \beta \frac{1+r}{c_{t+1}^i} + \Lambda_t^i(1+r) = 0.$

Complementary slackness conditions are given by (O3-4).

By rearranging (O1-4) and (O1-6), we have

(O1-7) $\Lambda_t^i = \beta \frac{1}{c_{t+1}^i} \frac{R^c - (1+r)}{1+r-\theta R^c}.$

Similarly, from (O1-5) and (O1-6), we have

(O1-8) $\Lambda_t^i = \beta \frac{1}{c_{t+1}^i} \frac{R_t^x - (1+r)}{1+r-\theta^x R_t^x}.$

Substituting (O1-7) into (O1-4) yields

(O1-9) $\frac{1}{c_t^i} = \beta \frac{1+1}{c_{t+1}^i} \frac{R^c(1-\theta)}{1+r-\theta R^c}.$

Similarly, substituting (O1-8) into (O1-6) yields

(O1-10) $\quad \frac{1}{c_t^i} = \beta \frac{1+1}{c_{t+1}^i} \frac{R_t^x(1-\theta^x)}{1+r-\theta^x R_t^x}.$

From (O1-9) and (O1-10), we obtain

(O1-11) $\quad \frac{R^c(1-\theta)}{1+r-\theta R^c} = \frac{R_t^x(1-\theta^x)}{1+r-\theta^x R_t^x},$

which is equivalent to the no-arbitrage condition (9), i.e.,

(9) $\quad \frac{R^c(1-\theta)}{1-\frac{\theta R^c}{1+r}} = \frac{R_t^x(1-\theta^x)}{1-\frac{\theta^x R_t^x}{1+r}}.$

Also, substituting (2) with the binding constraint into (O1-2a) and (O1-2b) yields

(O1-11a) $\quad c_t^i + (1 - \frac{\theta R^C}{1+r})k_{t+1}^i + (1 - \frac{\theta^x R_t^x}{1+r})P_t x_t^i = w_t.$

Considering (9) yields

(O1-11b) $\quad c_{t+1}^i = (R_{t+1} + 1 - \delta)(1-\theta)k_{t+1}^i + \left(\frac{D_{t+1}+P_{t+1}}{P_t}\right)(1-\theta^x)P_t x_t^i = \frac{R^c(1-\theta)}{1-\frac{\theta R^C}{1+r}}\left((1 - \frac{\theta R^C}{1+r})k_{t+1}^i + (1 - \frac{\theta^x R_t^x}{1+r})P_t x_t^i\right).$

By substituting (O1-11) into (O1-11b), we have

(O1-12) $\quad c_{t+1}^i = \frac{R^c(1-\theta)}{1-\frac{\theta R^C}{1+r}}(w_t - c_t^i).$

Then, by substituting (O1-12) into (O1-9) and solving for $c_t^i$, we have

(O1-13) $\quad c_t^i = \frac{1}{1+\beta} w_t$ and $w_t - c_t^i = \frac{\beta}{1+\beta} w_t.$

Since $\frac{\beta}{1+\beta}$ is the saving rate and constant, the analysis in the main text will apply by multiplying the saving rate to aggregate savings $\eta(1-\alpha)AK_t$.

## O2. Robustness: More general production functions for real estate sector

In this appendix, we extend the bare-bones model to the situation where both labor and land are inputs into real estate.

In the real estate sector, competitive firms produce the final consumption goods by using labor and land as factors for production. The aggregate production function is given by

(O2-1) $Y_t = D_t(N_t^X)^\rho (X)^{1-\rho}$,

where $N_t^X$ is labor input in the real estate sector.

The wage rates in the real estate sector and the productive sector are, respectively, given by

(O2-2) $w_t^X = \rho D_t (N_t^X)^{\rho-1}(X)^{1-\rho}$,

(O2-3) $w_t^K = (1-\alpha)A(N_t^K)^{-\alpha}K_t$,

where $N_t^K$ is labor input in the productive sector.

The labor market clearing condition is

(O2-4) $N_t^X + N_t^K = N = 1$.

We first consider a case where labor mobility is possible between the two sectors. Then, in equilibrium, the wage rates must be equal, i.e., $w_t^X = w_t^K \equiv w_t$. Considering $D_t = \epsilon a K_t$ and $X = 1$, we have

(O2-5) $(1-\alpha)A(N_t^K)^{-\alpha}K_t = \rho \epsilon a K_t (N_t^X)^{\rho-1}$,

which can be written as

(O2-6) $(1-\alpha)A(1-N_t^X)^{-\alpha} = \rho \epsilon a (N_t^X)^{\rho-1}$.

The left-hand side is a convex function of $N_t^X$ with a positive slope and a positive intercept $(1-\alpha)A$ at $N_t^X = 0$ and goes to infinity as $N_t^X$ goes to 1, while the right-hand side is also a convex function of $N_t^X$ with a negative slope and goes to infinity as $N_t^X$ goes to zero and becomes equal to $\rho \epsilon a$ at $N_t^X = 1$. Hence, $N_t^X$ is uniquely determined and is constant, and so is $N_t^K$.

Hence, total incomes of entrepreneurs from the two sectors can be written as

(O2-7) $\eta w_t^X N^X + \eta w_t^K N^K = \eta w_t = \eta(1-\alpha)A(N^K)^{-\alpha}K_t.$

Next, we examine a case where labor mobility is not possible. In this case, total wage incomes can be written as

(O2-8) $\eta w_t^X N^X + \eta w_t^K N^K = \eta[\rho\epsilon a(N^X)^{\rho-1} + (1-\alpha)A(N^K)^{-\alpha}]K_t.$

In (O2-7) and (O2-8), total incomes of entrepreneurs are a linear function of aggregate capital stock $K_t$. Hence, the analysis in the main text will apply by replacing $\eta w_t$ in (13) with (O2-7) or (O2-8).

## O3. Robustness: Imperfect spillovers

In the main text, to obtain the balanced growth path, we focused on the case where there is a full spillover from the productive sector to the real estate sector, so that productivities between the two sectors grow at the equal rate. In this Appendix, as robustness of our main analysis, we relax this assumption and examine a case where the two sectors grow at different rates. We will show that we obtain the same results as the ones in the main text.

We consider a case where the productivity growth rate of the real estate sector is given by

(O3-1) $D_{t+1} = (1+d)D_t$

i.e., productivity of the real estate sector will grow at the rate of $d \geq 0$. $d$ does not need to be the same as the growth rate of aggregate capital. Indeed, as we will analyze below, the two sectors grow at different rates, generating unbalanced growth.

By defining $n_t \equiv \frac{D_t}{AK_t}$ as the dividend share, the dynamics of this economy can be characterized by the following equations.

(O3-2) $\phi_{t+1} = \left(\frac{\frac{\lambda}{1-\theta^x + \frac{\theta^x \lambda}{1+r^*}}}{1+g_t}\right)\phi_t - \left(\frac{1+d}{1+g_t}\right)n_t.$

(O3-3) $n_{t+1} = \left(\frac{1+d}{1+g_t}\right)n_t.$

(O3-4) $1+g_t = \frac{A}{1-\frac{\theta R^c}{1+r}}\left[\eta(1-\alpha) - \left(\frac{1-\theta^x}{1-\theta^x+\frac{\theta^x\lambda}{1+r}}\right)\phi_t\right].$

Hence, the dynamical system can be written as

(O3-5) $\phi_{t+1} = f(\phi_t, n_t)$

and

(O3-6) $n_{t+1} = h(\phi_t, n_t)$,

where $n_t$ is a state variable and $\phi_t$ is a jump variable.

We impose conditions on parameter values.

**Assumption 2.** $\dfrac{\eta(1-\alpha)A}{1-\frac{\theta R^C}{1+r}} > R^x = \dfrac{\lambda}{1-\theta^x+\frac{\theta^x \lambda}{1+r}}$

**Assumption 5.** $R^x = \dfrac{\lambda}{1-\theta^x+\frac{\theta^x \lambda}{1+r}} > 1+d$

We study the asymptotic behaviour of this economy. Linearizing (O3-5) and (O3-6) in the neighbourhood of $(\phi^{**}, n^*) = \left( \dfrac{\frac{\eta(1-\alpha)A}{1-\frac{\theta R^C}{1+r}} - R^x}{\left(\frac{A}{1-\frac{\theta R^C}{1+r}}\right)\left(\frac{1-\theta^x}{1-\theta^x+\frac{\theta^x \lambda}{1+r}}\right)}, 0 \right)$, we obtain

(O3-7) $\begin{pmatrix} \phi_{t+1} - \phi^{**} \\ n_{t+1} - n^* \end{pmatrix} = \begin{pmatrix} \frac{\partial f(\phi^{**}, n^*)}{\partial \phi_t} & \frac{\partial f(\phi^{**}, n^*)}{\partial n_t} \\ \frac{\partial h(\phi^{**}, n^*)}{\partial \phi_t} & \frac{\partial h(\phi^{**}, n^*)}{\partial n_t} \end{pmatrix} \begin{pmatrix} \phi_t - \phi^{**} \\ n_t - n^* \end{pmatrix}$

with

(O3-8) $\dfrac{\partial f(\phi^{**}, n^*)}{\partial \phi_t} = \dfrac{\eta(1-\alpha)A}{1-\frac{\theta R^C}{1+r}} / R^x$.

(O3-9) $\dfrac{\partial f(\phi^{**}, n^*)}{\partial n_t} = -\dfrac{1+d}{R^x}$.

(O3-10) $\dfrac{\partial h(\phi^{**}, n^*)}{\partial \phi_t} = 0$.

(O3-11) $\dfrac{\partial h(\phi^{**}, n^*)}{\partial n_t} = \dfrac{1+d}{R^x}$.

Hence, we have

(O3-12) $Trace = \frac{\partial f(\phi^{**},n^*)}{\partial \phi_t} + \frac{\partial h(\phi^{**},n^*)}{\partial m_t} = \frac{\eta(1-\alpha)A}{1-\frac{\theta R^C}{1+r}}/R^x + \frac{1+d}{R^x}$.

(O3-13) $Determinant = \frac{\partial f(\phi^{**},n^*)}{\partial \phi_t}\frac{\partial h(\phi^{**},n^*)}{\partial m_t} - \frac{\partial f(\phi^{**},n^*)}{\partial n_t}\frac{\partial h(\phi^{**},n^*)}{\partial \phi_t} = \left(\frac{\eta(1-\alpha)A}{1-\frac{\theta R^C}{1+r}}/R^x\right)\left(\frac{1+d}{R^x}\right) > 0$.

Under Assumption 2 and 5,

(O3-14) $Determinant - (Trace - 1) = \left(\left(\frac{\eta(1-\alpha)A}{1-\frac{\theta R^C}{1+r}}/R^x\right) - 1\right)\left(\frac{1+d}{R^x}\right) < 0$

Since $Determinant > 0$ and $Determinant < Trace - 1$, the equilibrium path is locally determinate. That is, given a sufficiently small $n_0 = \frac{D_0}{AK_0}$ as an initial value, there is a unique dynamic path with unbalanced growth where the productive sector and the real estate sector grow at different rates, and the economy will asymptotically converge to $(\phi^{**}, n^*)$.

Along the unbalanced growth dynamics, we have

(O3-15) $\lim_{t\to\infty} \frac{P_{t+1}}{P_t} = \frac{K_{t+1}}{K_t}$ and $\lim_{t\to\infty} \frac{K_{t+1}}{K_t} = R^x$.

Since aggregate capital will grow faster than land rents, the dividend share $n_t$ will become asymptotically sufficiently small. Therefore,

(O3-16) as $t \to \infty$, $1 + g_t \sim R^x$.

That is, if we focus on the asymptotic behaviour of this economy, the growth rate of this economy can approximately become $R^x$. This is qualitatively the same as the case of $\epsilon \to 0$ in our main analysis because the dividend share can be written as $\frac{D_t}{AK_t} = \frac{\epsilon a K_t}{AK_t} = \epsilon(a)^\alpha$, and as $\epsilon \to 0$, the dividend share becomes sufficiently small. Hence, this unbalanced growth dynamics can be a micro-foundation for our analysis with $\epsilon \to 0$.

It should be noted that instead of (O3-1), we can, for instance, consider

(O3-17) $D_t = \epsilon\chi(K_t) = \epsilon a K_t^\psi$, with $0 < \psi < 1$.

When $\psi = 1$, (O3-17) is the same as (3b). If $0 < \psi < 1$, i.e., imperfect spillovers, capital grows faster than land rents. Hence, the asymptotic behavior of this economy is the same as the analyses studied above.

## O4. Equilibrium land prices vs. "Fundamental Values"

We can also study whether the equilibrium land price in our bare-bones model equals its fundamental value, i.e. the present discounted value of the land rents. To do so, we need to derive the no-arbitrage condition in a consistent manner with the individual's optimization problem, i.e., the first-order conditions, and then define the fundamental value of land.

In our model, the no-arbitrage condition is given by (9). When we solve the equation for $P_t$ at the balanced growth path, we have

(O4-1) $\quad P_t = \dfrac{\epsilon a K_{t+1}}{\frac{\lambda}{1-\theta^x+\frac{\theta^x \lambda}{1+r^*}}} + \dfrac{P_{t+1}}{\frac{\lambda}{1-\theta^x+\frac{\theta^x \lambda}{1+r^*}}}.$

Note that the price of land is the discounted "dividend" (rent) and the discounted value of the price at which the land can be sold next period, with the discount rate being equal to $\dfrac{\lambda}{1-\theta^x+\frac{\theta^x \lambda}{1+r^*}}$, not $1 + r$ because entrepreneurs invest in capital and buy land, both using leverage; and that $\dfrac{\lambda}{1-\theta^x+\frac{\theta^x \lambda}{1+r^*}} > 1 + r^*$ because $R^c > 1 + r^*$. The discount rate in (O4-1) includes the effect of land relaxing the borrowing constraints: From (9), if $\theta^x > \theta$, the collateral value of land is greater than that of capital investment, the unleveraged rate of return on land $R^x$ is strictly lower than that on capital investment $R^c$. Despite this, entrepreneurs want to buy land because its collateral value is greater, i.e., holding land has a collateral premium, i.e., the collateral premium can be written as $R^c - \dfrac{\lambda}{1-\theta^x+\frac{\theta^x \lambda}{1+r^*}}$. Other things being constant, a rise in $\theta^x$ increases the land collateral premium, raising the land price.

Iterating (O4-1) forward, we obtain

(O4-2) $\quad P_t = \dfrac{\epsilon a K_{t+1}}{\frac{\lambda}{1-\theta^x+\frac{\theta^x \lambda}{1+r^*}}} + \dfrac{\epsilon a K_{t+2}}{\left(\frac{\lambda}{1-\theta^x+\frac{\theta^x \lambda}{1+r^*}}\right)^2} + \dfrac{\epsilon a K_{t+3}}{\left(\frac{\lambda}{1-\theta^x+\frac{\theta^x \lambda}{1+r^*}}\right)^3} + \cdots + \dfrac{\epsilon a K_{t+n}}{\left(\frac{\lambda}{1-\theta^x+\frac{\theta^x \lambda}{1+r^*}}\right)^n} + \dfrac{P_{t+n}}{\left(\frac{\lambda}{1-\theta^x+\frac{\theta^x \lambda}{1+r^*}}\right)^n},$

where $K_{t+1} = (1 + g^*)K_t$. The current land price equals the discounted value of land rents stream and the discounted value of the land price in the future.

We say that the equilibrium land price equals its fundamental value $V_t$, i.e., $P_t = V_t$, if and only if

(O4-3) $\quad \lim_{n \to \infty} \dfrac{P_{t+n}}{\left(\dfrac{\lambda}{1-\theta^x + \dfrac{\theta^x \lambda}{1+r^*}}\right)^n} = 0$,

(O4-3) is called the transversality condition for land prices. The economic meaning of $\lim_{n \to \infty} \dfrac{P_{t+n}}{\left(\dfrac{\lambda}{1-\theta^x + \dfrac{\theta^x \lambda}{1+r^*}}\right)^n}$ is that it captures a purely speculative aspect, that is, one buy land now for the purpose of resale, rather than for the purpose of receiving dividends in the future. (O4-3) captures its impact on the current land prices.

It is easy to establish that (O4-3) is satisfied, because the price of land is increasing at the rate $g^*$, and $\dfrac{\lambda}{1-\theta^x + \dfrac{\theta^x \lambda}{1+r^*}} > 1 + g^*$ from (14c), it is obvious that (O4-3) holds.

When (O4-3) is satisfied, we can define the fundamental value as

(O4-4) $\quad V_t \equiv \sum_{n=1}^{\infty} \epsilon a K_{t+n} \Big/ \left(\dfrac{\lambda}{1-\theta^x + \dfrac{\theta^x \lambda}{1+r^*}}\right)^n$.

That is, land prices (O4-2, iterated infinitely) are determined only by future discounted dividends (as expressed in (O4-4)). In the following Appendix O5, we explore the possibility of land price bubbles.

## O5. The Possibility of land price bubbles

When we consider unbalanced growth dynamics explored in Appendix O3, the transversality condition (O4-3) is violated, generating land price bubbles.

From the no-arbitrage equation, the fundamental value of land is defined as

(O5-1) $\quad V_t = \dfrac{D_{t+1}}{\dfrac{\lambda}{1-\theta^x + \dfrac{\theta^x \lambda}{1+r^*}}} + \dfrac{D_{t+2}}{\left(\dfrac{\lambda}{1-\theta^x + \dfrac{\theta^x \lambda}{1+r^*}}\right)^2} + \dfrac{D_{t+3}}{\left(\dfrac{\lambda}{1-\theta^x + \dfrac{\theta^x \lambda}{1+r^*}}\right)^3} \cdots = \dfrac{(1+d)D_t}{\dfrac{\lambda}{1-\theta^x + \dfrac{\theta^x \lambda}{1+r^*}} - (1+d)}$.

That is, under Assumption 5, (O5-1) can be written as (O5-2) $\quad \frac{V_t}{D_t} = \frac{(1+d)}{\frac{\lambda}{1-\theta^x + \frac{\theta^x \lambda}{1+r^*}} - (1+d)} < \infty.$

On the other hand, we know that as $t \to \infty$, $\frac{P_{t+1}}{P_t} \sim R^x > \frac{V_{t+1}}{V_t} = 1 + d$. Therefore, $P_t > V_t$. Intuitively, along the unbalanced growth dynamics, land prices will grow at a faster rate than land rents, generating a land price bubble. Moreover, in the unbalanced growth dynamics, (O3-15) holds. That is, $\lim_{t \to \infty} \frac{P_{t+1}}{P_t} = R^x$. Hence, the transversality condition (O4-3) is violated, i.e., $\lim_{n \to \infty} \frac{P_{t+n}}{(\frac{\lambda}{1-\theta^x + \frac{\theta^x \lambda}{1+r^*}})^n} > 0$ and equilibrium land prices contain a purely speculative aspect.

Hirano and Toda (2023a, 2023b) and Hirano, Jinnai, and Toda (2023) provide a theoretical foundation for the connection between unbalanced growth dynamics and asset price bubbles.

**O6. Robustness:** *Weakening the assumptions of limited participation in the credit market and credit frictions*

We need to distinguish between the return on a safe asset (a government bond. If there is uncertainty about the rate of inflation, bonds that are not inflation protected are still risky.) and a return on a risky investment. Both land and capital are risky, and not surprisingly, on average yield substantially higher returns than government bonds. This is what we would expect in equilibrium. Credit frictions inevitably follow, not only from asymmetries of information, but also from limited liability (or more generally, even in the absence of limited liability, the limited ability to recover funds.)

Think, for a moment, of workers as infinitely risk averse. Then they would only be willing to lend a limited amount either to an entrepreneur buying capital or buying land. There is a natural limit on leverage, even in the absence of policy. Those limits on leverage will, of course, mean that even with diminishing returns in production, the average return on capital and on land will remain above the safe rate of interest. Similar results would hold in a model with highly risk averse workers.

The model we have formulated here captures in a simple way these credit frictions, though crucially, we have not explicitly modelled either asymmetries of information or risk aversion. But it explains clearly why, for instance, there can be an equilibrium with productive land having a finite price, even if $r$ is very low. The non-entrepreneurial savers see (rightly) land holdings as risky, and so even if $r$ were zero, they would only have a finite demand for it, and its price would be finite.

Similarly, if government were to issue an excessive number of bonds, capital accumulation would be limited, the return on capital would increase, and the private sector could issue an increasing number of safe bonds: there is thus a limit to the *equilibrium* issuance of bonds by government at low interest rates at an interest rate less than $g$. Thus, while it might *seem* that with low $r$, government debt *could* be unbounded, the interest rate is low only because government borrowing is limited.

Thus, what have been posed as critical conundrums in the literature are really a reflection of the oversimplifications employed in the models in which they arise and the attempt to interpret the real world's data through the lens of those models.